\documentclass[12pt]{article}

\usepackage{a4,a4wide}

\usepackage{amsmath}
\usepackage{amsfonts}
\usepackage{fancybox}
\usepackage[dvips]{color}

\usepackage{color}
\definecolor{MyDarkBlue}{rgb}{0.15,0.15,0.45}
\usepackage[linktocpage=true]{hyperref}
\hypersetup{
colorlinks=true,
citecolor=MyDarkBlue,
linkcolor=MyDarkBlue,
urlcolor=MyDarkBlue,
}

\def\tr{\mathrm{tr}}
\def\Tr{\mathrm{Tr}}
\def\STr{\mathrm{STr}}

\def\half{{1\over2}}
\def\nn{\nonumber\\}

\def\[{\left[}
\def\]{\right]}
\def\({\left(}
\def\){\right)}

\newcommand{\wt}{\widetilde}

\def\={\stackrel{\bullet}{=}}
\def\nn{\notag\\}

\def\Tr{\mathrm{Tr}}

\def\half{{1\over2}}

\def\[{\left[}
\def\]{\right]}
\def\({\left(}
\def\){\right)}

\def\cL{{\cal L}}
\def\cN{{\cal N}}

\def\bA{{\mathbf A}}
\def\bB{{\mathbf B}}

\def\bZ{{\mathbf Z}}

\def \be {\begin{equation}}
\def \ee {\end{equation}}
\def \bea {\begin{eqnarray}}
\def \eea {\end{eqnarray}}
\def\beal#1{\begin{align}#1\end{align}}
\def \nn {\notag\\}
\def\sla#1{\not\!\!#1}

\def\aver#1{\left\langle #1 \right\rangle}

\begin{document}

\makeatletter
\renewcommand{\theequation}{%
\thesection.\arabic{equation}}
\@addtoreset{equation}{section}
\makeatother

\begin{titlepage}
\vspace{-3cm}
\title{
\begin{flushright}
\normalsize{ TIFR/TH/13-26\\
Sep 2013}
\end{flushright}
       \vspace{1.5cm}
A Note on Large $N$ Thermal Free Energy in Supersymmetric Chern-Simons Vector Models
       \vspace{1.5cm}
}
\author{
Shuichi Yokoyama
\\[25pt] 
\!\!\!\!\!\!\!\!{\it \normalsize Department of Theoretical Physics, Tata Institute of Fundamental Research,}\\
\!\!\!\!\!\!\!\!{\it \normalsize Homi Bhabha Road, Mumbai 400005, India}\\
\\[10pt]
\!\!\!\!\!\!\!\!{\small \tt E-mail: yokoyama(at)theory.tifr.res.in}
}

\date{}

\maketitle

\thispagestyle{empty}

\vspace{.2cm}

\begin{abstract}
\vspace{0.3cm}
\normalsize

We compute the exact effective action for $\cN=3$ $U(N)_k$ and $\cN=4,6$ 
$U(N)_k\times U(N')_{-k}$ Chern-Simons theories with 
minimal matter content in the 't Hooft vector model limit under which 
$N$ and $k$ go to infinity holding $N/k, N'$ fixed. 
We also extend this calculation to $\cN=4,6$ mass deformed case.
We show that those large $N$ effective actions except mass-deformed $\cN=6$ case 
precisely reduce to that of $\cN=2$ $U(N)_k$ Chern-Simons theory 
with one fundamental chiral field up to overall multiple factor.  
By using this result we argue the thermal free energy and self-duality of 
the $\cN=3,4,6$ Chern-Simons theories including the $\cN=4$ mass term 
reduce to those of the $\cN=2$ case under the limit.

\end{abstract}

\end{titlepage}

% \tableofcontents

\section{Introduction}

Recently there has been a big progress in study of three dimensional 
vector models with Chern-Simons gauge interaction (Chern-Simons vector models). 
A key discovery which triggered this progress is that 
this class of quantum field theories turned out to conserve an infinite number of 
higher spin currents in the 't Hooft limit and be exactly solvable under the limit
with the light-cone gauge \cite{Giombi:2011kc,Aharony:2011jz}.  

It was shown by a general argument 
that in a three dimensional conformal field theory 
keeping the ``almost'' conserved higher spin currents 
the form of the three point functions of higher spin currents is determined 
up to two undetermined parameters 
\cite{Maldacena:2012sf}. (See also \cite{Giombi:2011rz,Maldacena:2011jn}.) 
In a conformal Chern-Simons vector model, 
these two parameters basically correspond to rank of gauge group and Chern-Simons level, 
and the precise parameter mapping was determined  
by explicit computation of several three point correlators 
for bosonic and fermionic Chern-Simons vector models   
\cite{Aharony:2012nh,GurAri:2012is}. 
These results strongly suggested 
that under the 't Hooft limit 
these conformally symmetric Chern-Simons vector models enjoy level-rank duality
known in pure Chern-Simons theories. 
(This suggestion was earlier made in \cite{Giombi:2011kc}.) 
This non-supersymmetric duality is reminiscent of bosonization in two dimensions because 
the duality transformation exchanges the conserved currents of bosonic Chern-Simons vector model  
and those of fermionic one.  

On the other hand, it is also possible that 
effective actions and thermal free energies in Chern-Simons vector models 
are computed exactly in the 't Hooft limit with the light-cone gauge 
\cite{Giombi:2011kc,Jain:2012qi,Oferb:2012}
including chemical potential and holonomy \cite{Yokoyama:2012fa}.
(Related works are \cite{Shenker:2011zf,Banerjee:2012gh}). 
It was shown that thermal free energies of Chern-Simons vector models 
also exhibit the non-supersymmetric duality mentioned above 
and supersymmetric duality known as Giveon-Kutasov (or Seiberg-like) duality \cite{Giveon:2008zn,Benini:2011mf}
by incorporating holonomy distribution obeying fermionic statistics in a high temperature limit \cite{Aharony:2012ns}. 
This feature of holonomy was observed earlier in the study of Chern-Simons theory 
on the space of two torus by using the canonical formalism \cite{Douglas:1994ex}.
The origin of this peculiar holonomy distribution at a high temperature 
was clarified from the standpoint of path integral formalism in the study of 
Chern-Simons vector models on two sphere with thermal time \cite{Jain:2013py}. 
It turned out that due to the fermionic holonomy distribution 
Chern-Simons vector models enjoy a novel thermal phase structure consistent with the duality transformation  \cite{Jain:2013py,Takimi:2013zca}.

The three dimensional duality can be extended to 
the most general renormalizable Chern-Simons vector models with one fundamental scalar and fermion 
under the 't Hooft limit \cite{Jain:2013gza}. 
This generalization enabled one to connect the known and unknown dualities
by taking certain massless limits or scaling limits. 
As a result, strong evidence was provided for the expectation that $\cN=1,2$ 
Giveon-Kutasov duality and non-supersymmetric duality in the Chern-Simons vector models 
can be connected by renormalization group flow.

In this paper we explore the duality structure in Chern-Simons matter theories with 
higher supersymmetry by adding more fields. 
Especially to achieve $\cN\geq4$ supersymmetry it is required to 
consider a non-simple gauge group such as $U(N)_k\times U(N')_{-k}$ 
\cite{Schwarz:2004yj,Gaiotto:2008sd}. 
We study a Chern-Simons system with such a gauge group by taking the limit 
defined by $N,k\to\infty$ with $N/k, N'$ fixed to reduce a system to a vector model.   

The rest of this paper is organized as follows. 
In Section \ref{effectiveaction},  
we compute the exact effective action of $\cN=3$ $U(N)_k$ and $\cN=4,6$ 
$U(N)_k\times U(N')_{-k}$ Chern-Simons theories including $\cN=4,6$ mass terms 
by taking the 't Hooft vector model limit.
In Section \ref{comments}, 
using the result obtained in Section \ref{effectiveaction}, 
we discuss the thermal free energy and self-duality of the supersymmetric Chern-Simons matter theories.%
\footnote{When we say one theory is self-dual, we mean that 
the dual theory thereof is in the same theory space with different parameters. }
Section \ref{discussion} is devoted to summary and discussion. 
In Appendix, supersymmetric Chern-Simons matter actions are written
in our convention. 

%%%%%%%%%%%%%%%%%%%%%%%%%%%%%%%%%%%%%%%
\section{Exact large $N$ effective action} 
\label{effectiveaction}
\subsection{Fundamental matter fields}

In this preliminary section we study $U(N)_k$ Chern-Simons theory 
coupling to $M$ fundamental scalar and fermionic fields in the 't Hooft limit, 
in which $N$ and  $k$ go to infinity with $\lambda=N/k$ fixed.
We denote $M$ fundamental scalar fields and fermionic ones by $q^A, \psi_A$ respectively, 
where $A=1,2, \cdots, M$. 
The case $M=1$ was studied in detail in \cite{Jain:2012qi}.
We are interested in a situation where the theory has $U(M)$ flavor symmetry, 
which we assume in what follows. 
The main purpose of this section is 
to demonstrate how to generalize the previous result to $M$ copies of the matter fields
reviewing the technique employed in the previous study.

For this purpose let us start by a generic action
\beal{
S = \int d^3 x ( \kappa \cL_{cs}[A] + \cL_{m}) .
\label{action}
}
Here $\kappa$ is related to the Chern-Simons level $k$ by $\kappa={k\over 4\pi}$ and 
\beal{
\cL_{cs}[A]  =& \Tr \biggl[ i \varepsilon^{\mu\nu\rho} \left(A_\mu\partial_\nu A_\rho -{2i\over3}  A_\mu  A_\nu A_\rho \right)\biggl], 
\label{csaction} \\
\cL_{m}=& D_\mu q^\dagger_A D^\mu q^A + \psi^\dagger{}^{  A} \gamma^\mu D_\mu \psi_{ A}
+ V_{m},
\label{fmatterlagrangian}
}
where $D_\mu$ is the covariant derivative acting on the fields in a way that 
\beal{
D_\mu q^A& = \partial_\mu q^A -i A_\mu q^A, \quad
D_\mu q^\dagger_A= \partial_\mu q^\dagger_A +i q^\dagger_A  A_\mu,\nn
D_\mu \psi_A&= \partial_\mu\psi_A -i A_\mu \psi_A,\quad
D_\mu \psi^\dagger{}^{A}= \partial_\mu\psi^\dagger{}^{A} +i \psi^\dagger{}^{A} A_\mu. 
\label{fcovderiv}
}
$V_m$ represents a gauge-invariant potential in this system 
given by a function of bilinears of the elementary fields $q^A, \psi_B$ in a flavor-singlet way. 
% \be
% V_{m}= V_{m}(q^\dagger_A q^B, \psi^\dagger{}^A \psi_B, q^\dagger_A \psi_B, \psi^\dagger{}^A q^B) 
% \ee
We suppress contraction of fundamental gauge indices for notational simplification. 
A specific example is $\cN=3$, whose action in our notation is given in \ref{n3action}.

Firstly we separate the gauge field into $U(1)$ part and $SU(N)$ one. 
The Chern-Simons coupling of $U(1)$ gauge field is given by $N k$, 
which means the gauge propagator of $U(1)$ gauge field has extra $1/N$ factor compared to 
that of $SU(N)$ part. 
Therefore the contribution of $U(1)$ part of the gauge field is sub-leading in the large $N$ limit.
% In other words, the large $N$ thermal free energy with $U(N)$ gauge group is the same as that of 
% $SU(N)$ gauge group in the 't Hooft limit. 

So let us focus on the case when the gauge group is $SU(N)$. 
In order to determine the exact effective action 
we fix the gauge degrees of freedom by the (Euclidean) light-cone gauge \cite{Giombi:2011kc}.  
This gauge fixing gets rid of the cubic interaction of gauge field, 
which enables us to integrate it out. 
From the equation of motion for $A_+$ we obtain 
\be
2\kappa \partial_- A^a_3 = \Tr\[ i\(  q^A \partial_- q^\dagger_A - \partial_- q^A q^\dagger _A
-  \psi_A \gamma_- \psi^\dagger{}^A\) T^a \]
\label{eoma+}
\ee
where $T^a$ is a generator of $SU(N)$ gauge group.
%and we suppress fundamental gauge indices contracted in the right-hand side. 
A solution in the Fourier space is given by%
\footnote{In this solution we neglect zero mode (holonomy) of this gauge field. 
Here we suppress this for notational simplification. 
The holonomy can be taken into account by shifting the momentum in the propagator \cite{Yokoyama:2012fa}. 
}
\be
A^a_3(q) = {1\over 2\kappa i q_-} \int {d^3 r \over (2\pi)^3} \Tr\[\( (2r+q)_- q^A(r+q) q^\dagger_A(-r)
+i  \psi_A(q+r) \gamma_- \psi^\dagger{}^A(-r) \) T^a\].
\label{solutiona3}
\ee
Plugging the solution into the action \eqref{action}, 
we find \cite{Jain:2012qi}
\beal{
S=& \int {d^3 p \over (2\pi)^3} ( p^2 q^\dagger_A(-p) q^A(p) 
+ \psi^\dagger{}^{  A}(-p) i \gamma^\mu p_\mu \psi_{ A}(p)) + S_{m}  \nn
&+ N \int  \frac{d^3P}{(2 \pi)^3} \frac{d^3q_1}{(2 \pi)^3} \frac{d^3q_2}{(2 \pi)^3} 
 C_1(P,q_1,q_2)  \chi^A_B(P,q_1) \chi^B_A(-P,q_2) \nn
&+ N \int \frac{d^3P_1}{(2 \pi)^3} \frac{d^3P_2}{(2 \pi)^3}  \frac{d^3q_1}{(2 \pi)^3} \frac{d^3q_2}{(2 \pi)^3}\frac{d^3q_3}{(2 \pi)^3} C_2(P_1,P_2,q_1,q_2,q_3)
\chi^A_B(P_1,q_1)\chi^B_C(P_2,q_2)\chi^C_A(-P_1-P_2,q_3)\nn
&+N \int \frac{d^3P}{(2 \pi)^3} \frac{d^3q_1}{(2 \pi)^3} \frac{d^3q_2}{(2 \pi)^3}~
\frac{8 \pi iN}{ k (q_1-q_2)_{-}} \xi^A_B{}_-(P,q_1) \xi^B_A{}_I(-P,q_2)
+ \cdots,
\label{s1}
}
where   
\beal{
&\chi^B_A(P,q)=\frac{1}{N}q^\dagger_A(\frac{P}{2}-q) q^B(\frac{P}{2}+q), \quad
\label{chi}\\
&\xi^B_A{}_I(P,q) = \frac{1}{2N}{\psi^\dagger}{}^B (\frac{P}{2}-q) \psi_A(\frac{P}{2}+q), \\
&\xi^B_A{}_-(P,q) = \frac{1}{2N}{\psi^\dagger}{}^B(\frac{P}{2}-q) \gamma_-\psi_A(\frac{P}{2}+q),
\label{xi-}\\
&C_1(P,q_1,q_2)={2 \pi i N \over k} \frac{(-P+q_1+ q_2)_{3} (P+q_1+q_2)_{-}}{(q_1-q_2)_{-}}, \\
&C_2(P_1,P_2,q_1,q_2,q_3)=
{4 \pi^2 N^2 \over k^2} \frac{( P_1 - P_2+2q_1+2q_2)_{-} (P_1+2P_2+ 2q_2 + 2q_3)_{-}}{(P_1+P_2+ 2q_1-2q_2)_{-} (P_1-2q_2+2q_3)_{-}}.
}
$S_m$ is 
\be
S_m = \int d^3 x V_m = S_m (\chi^B_A, \xi^A_B{}_-, \xi^A_B{}_I, \eta_{AB}, \bar\eta^{AB}) 
\label{sm}
\ee
where
\be
\eta_{AB}(P,q)= \frac{1}{N}q^\dagger_A(\frac{P}{2}-q) \psi_B(\frac{P}{2}+q) , \quad 
\bar\eta^{AB}(P,q)= \frac{1}{N}\psi^\dagger{}^A(\frac{P}{2}-q) q^B(\frac{P}{2}+q).
\ee
The ellipsis in \eqref{s1} represents $1/N$ correction terms and those which contain $\eta_{AB}, \bar\eta^{AB}$.

The next step is to introduce auxiliary bilocal fields so that the interaction terms disappear in the action. 
We can add the following terms without changing the dynamics
\beal{
\Delta S=& - N\int \frac{d^3 P}{(2 \pi)^3}\frac{d^3 q}{(2 \pi)^3}
\biggl( \Sigma^A_B(P, q) (\alpha^B_A(-P,q) - \chi^B_A(-P, q)) \nn 
&+  2\Pi_B^A{}^I(P, q) (\beta^B_A{}_I(-P,q) - \xi^B_A{}_I(-P,q)) 
+ 2\Pi_B^A{}^-(P, q) (\beta^B_A{}_-(-P,q) - \xi^B_A{}_-(-P,q))\nn 
&+\Gamma^{AB}(P,q)(\gamma_{AB}(-P,q) -  \eta_{AB}(-P,q)) 
+\bar\Gamma_{AB}(P,q)(\bar\gamma^{AB}(-P,q) - \bar \eta^{AB}(-P,q)) \biggl) \nn
&-S_{int}(\chi^B_A, \xi^B_A{}_I, \xi^B_A{}_-,\eta_{AB},\bar\eta^{AB}) 
+ S_{int}(\alpha^B_A, \beta^B_A{}_I, \beta^B_A{}_-,\gamma_{AB},\bar\gamma^{AB}) 
\label{deltaS}
}
where $S_{int}(\chi^B_A, \xi^B_A{}_I, \xi^B_A{}_-,\eta_{AB},\bar\eta^{AB}) $ denotes 
all interaction terms in \eqref{s1}. 
Adding this into \eqref{s1} gives 
\beal{
&S+\Delta S\nn
=&\int \frac{d^3P}{(2 \pi)^3} \frac{d^3q}{(2 \pi)^3} \bigl(q^\dagger_B({P\over 2}-q), \psi^\dagger{}^B({P\over 2}-q)\bigl)
\boldsymbol Q
\left(\begin{array}{c} q^A({P\over2}+q) \\ \psi_A({P\over2}+q) \end{array}\right) + S_{m} (\alpha^B_A, \beta^B_A{}_I, \beta^B_A{}_-,\gamma_{AB},\bar\gamma^{AB})  \nn
&+ N \int  \frac{d^3P}{(2 \pi)^3} \frac{d^3q_1}{(2 \pi)^3} \frac{d^3q_2}{(2 \pi)^3} 
 C_1(P,q_1,q_2)  \alpha^A_B(P,q_1) \alpha^B_A(-P,q_2) \nn
&+ N \int \frac{d^3P_1}{(2 \pi)^3} \frac{d^3P_2}{(2 \pi)^3}  \frac{d^3q_1}{(2 \pi)^3} \frac{d^3q_2}{(2 \pi)^3}\frac{d^3q_3}{(2 \pi)^3} C_2(P_1,P_2,q_1,q_2,q_3)
\alpha^A_B(P_1,q_1)\alpha^B_C(P_2,q_2)\alpha^C_A(-P_1-P_2,q_3)\nn
&+N \int \frac{d^3P}{(2 \pi)^3} \frac{d^3q_1}{(2 \pi)^3} \frac{d^3q_2}{(2 \pi)^3}~
\frac{8 \pi iN}{ k (q_1-q_2)_{-}} \beta^A_B{}_-(P,q_1) \beta^B_A{}_I(-P,q_2)\nn
&- N\int \frac{d^3 P}{(2 \pi)^3}\frac{d^3 q}{(2 \pi)^3}
\biggl( \Sigma^A_B(P, q) \alpha^B_A(-P,q) +  2\Pi_B^A{}^I(P, q) \beta^B_A{}_I(-P,q) 
+ 2\Pi_B^A{}^-(P, q) \beta^B_A{}_-(-P,q) \nn 
&~~~~~~~~~~~~~~~+\Gamma^{AB}(P,q)\gamma_{AB}(-P,q)  
+\bar\Gamma_{AB}(P,q)\bar\gamma^{AB}(-P,q) \biggl) 
+ \cdots,
\label{s2}
}
where 
\bea 
\boldsymbol Q
= \left(
\begin{array}{cc}
q^2 \delta^3(P)\delta^A_B + \Sigma^A_B(P,q) & \Gamma_{AB}(P,q) \\
\bar\Gamma^{AB}(P,q) & i \gamma^\mu q_\mu\delta^3(P)\delta^A_B + \Pi^A_B(P,q)
\end{array}\right),
\eea
and the ellipsis contains $1/N$ sub-leading and $\gamma_{AB}, \bar\gamma^{AB}$ terms. 
Since this is quadratic in terms of the elementary fields $q^A, \psi_A$, 
they are integrated out by gaussian integration, which results in 
\beal{
S_{\text{eff}}=&\STr \log \boldsymbol Q + S_{m} (\alpha^B_A, \beta^B_A{}_I, \beta^B_A{}_-,\gamma_{AB},\bar\gamma^{AB})  \nn
&+ N \int  \frac{d^3P}{(2 \pi)^3} \frac{d^3q_1}{(2 \pi)^3} \frac{d^3q_2}{(2 \pi)^3} 
 C_1(P,q_1,q_2)  \alpha^A_B(P,q_1) \alpha^B_A(-P,q_2) \nn
&+ N \int \frac{d^3P_1}{(2 \pi)^3} \frac{d^3P_2}{(2 \pi)^3}  \frac{d^3q_1}{(2 \pi)^3} \frac{d^3q_2}{(2 \pi)^3}\frac{d^3q_3}{(2 \pi)^3} C_2(P_1,P_2,q_1,q_2,q_3)
\alpha^A_B(P_1,q_1)\alpha^B_C(P_2,q_2)\alpha^C_A(-P_1-P_2,q_3)\nn
&+N \int \frac{d^3P}{(2 \pi)^3} \frac{d^3q_1}{(2 \pi)^3} \frac{d^3q_2}{(2 \pi)^3}~
\frac{8 \pi iN}{ k (q_1-q_2)_{-}} \beta^A_B{}_-(P,q_1) \beta^B_A{}_I(-P,q_2)\nn
&- N\int \frac{d^3 P}{(2 \pi)^3}\frac{d^3 q}{(2 \pi)^3}
\biggl( \Sigma^A_B(P, q) \alpha^B_A(-P,q) +  2\Pi_B^A{}^I(P, q) \beta^B_A{}_I(-P,q) 
+ 2\Pi_B^A{}^-(P, q) \beta^B_A{}_-(-P,q) \nn 
&~~~~~~~~~~~~~~~
+\Gamma^{AB}(P,q)\gamma_{AB}(-P,q)  
+\bar\Gamma_{AB}(P,q)\bar\gamma^{AB}(-P,q) \biggl) 
+ \cdots.
\label{s3}
}

Our interest is in the leading behavior of the large $N$ limit. 
For this purpose we shall focus on evaluating this on the saddle point. 
A natural ansatz for saddle point equations is such that 
solutions satisfy the translational, rotational invariance 
and covariance with respect to flavor indices. 
\beal{
&\aver{\alpha^A_B(P,q)}=(2 \pi)^3 \delta^3(P) \delta^A_B\alpha(q), \;
\aver{\beta^A_B(P,q)}= (2 \pi)^3 \delta^3(P) \delta^A_B\beta(q),\;
\aver{\gamma_{AB} (P,q)}=\aver{\bar\gamma^{AB} (P,q)}= 0, \nn
&\aver{\Sigma^A_B(P,q)}=(2 \pi)^3 \delta^3(P) \delta^A_B\Sigma(q), \;
\aver{\Pi^A_B(P,q)}= (2 \pi)^3 \delta^3(P) \delta^A_B\Pi(q),\;
\aver{\Gamma_{AB} (P,q)}=\aver{\bar\Gamma^{AB} (P,q)}= 0.
 \label{assumptions}
 }
Under this assumption the above effective action in the leading of large $N$ is simplified to be
\beal{
S_{\text{eff}}=&NMV \biggl[ \int \frac{d^3q}{(2 \pi)^3} \left( \log (q^2 +\Sigma(q) )
-\tr\log ( i \gamma^\mu q_\mu + \Pi(q)) \right) \nn
&+\int \frac{d^3q_1}{(2 \pi)^3} \frac{d^3q_2}{(2 \pi)^3} 
 C_1(q_1,q_2)  \alpha(q_1) \alpha(q_2) \nn
&+ \int \frac{d^3q_1}{(2 \pi)^3} \frac{d^3q_2}{(2 \pi)^3}\frac{d^3q_3}{(2 \pi)^3} C_2(q_1,q_2,q_3)
\alpha(q_1)\alpha(q_2)\alpha(q_3)\nn
&+ \int \frac{d^3q_1}{(2 \pi)^3} \frac{d^3q_2}{(2 \pi)^3}~
\frac{8 \pi iN}{ k (q_1-q_2)_{-}} \beta{}_-(q_1) \beta{}_I(q_2)\nn
&- \int \frac{d^3 q}{(2 \pi)^3}
\biggl( \Sigma(q) \alpha(q) +  2\Pi{}^I(q) \beta{}_I(q) 
+ 2\Pi^-(q) \beta_-(q) \biggl) \biggl]\nn
&+ S_{m} (\alpha\delta^B_A, \beta{}_I\delta^B_A, \beta{}_-\delta^B_A,0,0),
\label{s4}
}
where $V=(2\pi)^3 \delta^3(P=0)$ and 
\beal{
C_1(q_1,q_2) = & \half (C_1(0,q_1,q_2) + C_1(0,q_2,q_1)), \\
 C_2(q_1,q_2,q_3) =& {1 \over 3!} ( C_2(0,0,q_1,q_2,q_3) + (\text{permutation}) ).
}
Note that the terms in the bracket are the same as those obtained in the same procedure 
from Chern-Simons theory with one fundamental boson and fermion ($M=1$). 

To proceed further we need to specify a potential form of matter fields. 
We shall do case study by using $\cN=3$ Chern-Simons theory in the next subsection.

\subsubsection{$\cN=3$ $U(N)_k$ case} 
\label{n3}

In this subsection we apply the result obtained in the previous section 
to $\cN=3$ $U(N)_k$ Chern-Simons theory with minimal matter content.
The matter content is two fundamental complex scalar fields $q^A$ and fermionic fields $\psi_A$, 
where $A=1,2$.
% $SU(N)_k$ case is the same as $U(N)$ case in the 't Hooft limit as mentioned in the previous section.

$\cN=3$ $U(N)_k$ Chern-Simons Lagrangian is given by \eqref{n3ukaction}. 
The potential of the matter fields reads from \eqref{n3ukaction} as follows. 
\beal{
V^{\cN=3}_m=& {1 \over  \kappa}(\psi^\dagger{}^{  A} \psi_B) (q^\dagger_A q^{  B}) 
+ {1 \over  \kappa}(\psi^\dagger{}^{  A} q^B) (q^\dagger_A \psi_{  B}) 
- {1 \over 2 \kappa} (\psi^\dagger{}^{  A} q^B) (q^\dagger_B \psi_{A}) \nn
&+ {1 \over 2 \kappa}  \varepsilon_{AB} \varepsilon_{CD} (\psi^\dagger{}^A q^B) 
(\psi^\dagger{}^C q^D) 
+ {1 \over 2\kappa}  \varepsilon^{AB} \varepsilon^{CD} ( q^\dagger_A\psi_B) (q^\dagger_C\psi_D) 
\nn
& + {1 \over \kappa^2}
\({1\over 3} (q_A^\dagger  q^B) (q^\dagger_C q^A) (q^\dagger_B q^C)
-{1\over 12} (q_A^\dagger  q^B) (q^\dagger_B q^C) (q^\dagger_C q^A)\).
}
We again contract gauge indices by bracket notation. For example,  
$(\bar  q^A  \psi_B)= \bar  q^A_m  \psi_B^m$, where $m$ is 
a gauge index of the fundamental representation. 
Therefore $S_m$ in \eqref{sm} is given by 
\beal{
&S^{\cN=3}_m (\chi^B_A, \xi^A_B{}_-, \xi^A_B{}_I, \eta_{AB}, \bar\eta^{AB})  \nn
=&N \int  \frac{d^3P}{(2 \pi)^3} \frac{d^3q_1}{(2 \pi)^3} \frac{d^3q_2}{(2 \pi)^3} 
 \biggl[{2N\over \kappa} \xi^A_B{}_I(P,q_1) \chi^B_A(-P,q_2) \nn
& + {N\over \kappa} \bar\eta^{AB}(P,q_1) \eta_{AB}(-P,q_2) 
- {N\over 2\kappa} \bar\eta^{AB}(P,q_1) \eta_{BA}(-P,q_2) \nn
& + {N \over 2 \kappa}  \varepsilon_{AB} \varepsilon_{CD}\bar\eta^{AB} (P,q_1)  \bar\eta^{CD} (-P,q_2) 
+ {N \over 2\kappa}  \varepsilon^{AB} \varepsilon^{CD} \eta_{AB} (P,q_1)  \eta_{CD} (-P,q_2) \biggl]
\nn
&+ N \int \frac{d^3P_1}{(2 \pi)^3} \frac{d^3P_2}{(2 \pi)^3}  \frac{d^3q_1}{(2 \pi)^3} \frac{d^3q_2}{(2 \pi)^3}\frac{d^3q_3}{(2 \pi)^3} {N^2 \over \kappa^2}
\biggl[ {1\over 3} \chi^B_A(P_1,q_1)\chi^A_C(P_2,q_2)\chi^C_B(-P_1-P_2,q_3)\nn
&-{1\over 12} \chi^B_A(P_1,q_1)\chi^C_B(P_2,q_2)\chi^A_C(-P_1-P_2,q_3) \biggl].
}
Under the assumption \eqref{assumptions} this is simplified as follows. 
\beal{
&S^{\cN=3}_{m} (\alpha\delta^B_A, \beta{}_I\delta^B_A, \beta{}_-\delta^B_A,0,0) \nn
=&2NV \[ \int \frac{d^3q_1}{(2 \pi)^3} \frac{d^3q_2}{(2 \pi)^3} 
 {2N\over \kappa} \beta_I(q_1) \alpha(q_2) + \int \frac{d^3q_1}{(2 \pi)^3} \frac{d^3q_2}{(2 \pi)^3}\frac{d^3q_3}{(2 \pi)^3} 
 {N^2 \over (2\kappa)^2} \alpha(q_1)\alpha(q_2)\alpha(q_3)\].
}

One will soon notice that this is twice as that of the $\cN=2$ $U(N)_k$ Chern-Simons theory with minimal matter content:
\be
S^{\cN=3}_{m} (\alpha\delta^B_A, \beta{}_I\delta^B_A, \beta{}_-\delta^B_A,0,0)
=2 S^{\cN=2}_{m} (\alpha, \beta{}_I, \beta{}_-,0,0).
\label{n3n2}
\ee
To show this, let us read off the matter potential in $\cN=2$ case from \eqref{n2ukaction}. 
\beal{
V^{\cN=2}_m=
{1 \over 2 \kappa} (\bar\psi q)( \bar q \psi)
+ {1 \over \kappa} (\bar\psi \psi) (\bar q q) 
+ \left( {1 \over 2 \kappa}\right)^2 (\bar q q)^3. 
}
In the same way, we can compute $S_m$
\beal{
&S^{\cN=2}_m (\chi, \xi_-, \xi_I, \eta, \bar\eta)  \nn
=&N \int  \frac{d^3P}{(2 \pi)^3} \frac{d^3q_1}{(2 \pi)^3} \frac{d^3q_2}{(2 \pi)^3} 
 \biggl[{2N\over \kappa} \xi{}_I(P,q_1) \chi(-P,q_2) + {N\over 2 \kappa} \bar\eta(P,q_1) \eta(-P,q_2) \biggl] \nn
&+ N \int \frac{d^3P_1}{(2 \pi)^3} \frac{d^3P_2}{(2 \pi)^3}  \frac{d^3q_1}{(2 \pi)^3} \frac{d^3q_2}{(2 \pi)^3}\frac{d^3q_3}{(2 \pi)^3} {N^2 \over (2\kappa)^2}\chi(P_1,q_1)\chi(P_2,q_2)\chi(-P_1-P_2,q_3),
}
and under the assumption \eqref{assumptions}, 
\beal{
&S^{\cN=2}_m (\alpha, \beta_-, \beta_I, 0, 0)  \nn
=&NV \[\int \frac{d^3q_1}{(2 \pi)^3} \frac{d^3q_2}{(2 \pi)^3} {2N\over \kappa} \beta{}_I(q_1) \alpha(q_2)  
 + \int \frac{d^3q_1}{(2 \pi)^3} \frac{d^3q_2}{(2 \pi)^3}\frac{d^3q_3}{(2 \pi)^3} {N^2 \over (2\kappa)^2}\alpha(q_1)\alpha(q_2)\alpha(q_3) \],
}
which proves the relation \eqref{n3n2}.

By taking account of \eqref{s4}, the total large $N$ effective action in the minimal $\cN=3$ Chern-Simons theory 
is exactly twice as that of the minimal $\cN=2$ Chern-Simons theory in the 't Hooft limit. 

One might wonder why the large $N$ effective action is insensitive to 
the difference between the $\cN=2$ Chern-Simons theory and $\cN=3$ one. 
To understand this, let us consider 
$\cN=2$ Chern-Simons theory with one pair of chiral/anti-chiral fields $(Q, \wt Q)$ 
perturbed by a superpotential of the form $W_0= a (\wt Q T^b Q)^2$, 
where $a$ is a small positive number. 
It was shown in \cite{Gaiotto:2007qi} that 
this $\cN=2$ Chern-Simons matter theory with the superpotential flows to 
the same $\cN=2$ one in the infra-red (IR) except that the superpotential is given by
$W = a_{IR} (\wt Q T^b Q)^2$, where $a_{IR}$ is a fixed number of order $1/\kappa$, and 
$\cN=2$ supersymmetry is enhanced to $\cN=3$ in the IR so that 
the IR theory becomes the same as the $\cN=3$ Chern-Simons theory considered above.    
On the other hand, in the large $N$ limit, large $N$ factorization occurs so that 
the leading contribution of the superpotential is given by 
$\aver{W} =  a_{IR} \langle \wt Q Q \rangle^2$, which vanishes 
on the $SU(2)$ symmetric vacuum \eqref{assumptions}.%
\footnote{We used $\sum_{b=1}^N  (T^b)_m^n (T^b)_p^q= 
\delta_m^q \delta_p^n$ to rewrite the form of superpotential.
} 
This is the reason why the large $N$ effective action cannot see the difference between
the $\cN=3$ Chern-Simons theory and the $\cN=2$ one with the same matter content.

%%%%%%%%%%%%%%%%%%%%%%%%%%%%%%%%%%%%%%%%%%%%
\subsection{Bi-fundamental matter fields}
\label{bifund}

In this section we study $U(N)_k\times U(N')_{-k}$ Chern-Simons theory coupling to $M$ bi-fundamental matter fields by taking $N$ and  $k$ to infinity and holding $\lambda=N/k$ and $N'$ fixed. 
We denote $M$ bi-fundamental scalar fields and fermions by $q^A, \psi_A$ respectively, where $A=1,2, \cdots, M$. 
A generic form of the action of this class of Chern-Simons theories is given by 
\beal{
S = \int d^3 x ( \kappa (\cL_{cs}[A] - \cL_{cs}[A']) + \cL_{m}),
\label{actionbif}
}
where $\cL_{m}$ is given by
\beal{
\cL_{m}=& \Tr[D_\mu q^\dagger_A D^\mu q^A + \psi^\dagger{}^{  A} \gamma^\mu D_\mu \psi_{ A}] 
+V_m. 
\label{bifmatterlagrangian}
}
Here the covariant derivative acts on the fields in a way that 
\beal{
D_\mu q^A& = \partial_\mu q^A -i A_\mu q^A +i q^A  A' {}_\mu, \quad
D_\mu q^\dagger_A= \partial_\mu q^\dagger_A -i A' {}_\mu q^\dagger_A +i q^\dagger_A  A_\mu,\nn
D_\mu \psi_A&= \partial_\mu\psi_A -i A_\mu \psi_A +i \psi_A A' {}_\mu,\quad
D_\mu \psi^\dagger{}^{A}= \partial_\mu\psi^\dagger{}^{A} -i A' {}_\mu \psi^\dagger{}^{A}+i
\psi^\dagger{}^{A} A_\mu. 
\label{bifcovderiv}
}
$V_m$ represents a gauge-invariant potential of the matter fields in this system.
Specific examples are $\cN=4$ and $6$ Chern-Simons-matter theories, 
whose actions in our notation are given in \ref{n4} and \ref{n6}. 

Firstly we separate the $U(1)$ gauge fields. 
\be
A_\mu \to b_\mu + A_\mu, \quad
A'_\mu \to b'_\mu + A'_\mu, \quad
\label{separation}
\ee
where $b_\mu, b'_\mu$ are the trace part and $A_\mu, A'_\mu$ are the traceless part.
After this replacement $A_\mu, A'_\mu$ always represent $SU(N), SU(N')$ gauge fields. 
Plugging this into the matter action gives 
\beal{
\cL_{m}\to& \Tr[(b^-_\mu)^2 q^\dagger_A q^A + i b^-_\mu(q^\dagger_A D^\mu q^A - D^\mu q^\dagger _A q^A - \psi^\dagger{}^A \gamma^\mu \psi_A)]+\cL_{m}
}
where $b_\mu^-:=b_\mu-b'_\mu$ and $\cL_{m}$ in the right-hand side is the same as \eqref{bifmatterlagrangian} except $A_\mu, A'_\mu$ are now $SU(N)$ and $SU(N')$ gauge fields. 
$D^\mu$ is the covariant derivative of the $SU(N)\times SU(N')$ gauge group. 
Only a relative combination of $U(1)\times U(1)$ gauge fields, $b^-_\mu$, couples to the matter fields.

Let us turn to Chern-Simons term and also separate the $U(1)$ part of the gauge fields from 
the Chern-Simons term. Substituting \eqref{separation} into the Chern-Simons term we obtain
\beal{
 \kappa (\cL_{cs}[A] - \cL_{cs}[A']) \to&   i \kappa \varepsilon^{\mu\nu\rho} {N (N-N') \over (\sqrt N + \sqrt N')^2} \left(b^+_\mu\partial_\nu b^+_\rho +{2 (\sqrt{N'N} + N') \over N-N'}  b^+_\mu  \partial_\nu b^-_\rho \right) \nn
&{}+\kappa (\cL_{cs}[A] - \cL_{cs}[A']), 
\label{csterm2}
}
where we define 
\be
b^+_\mu = b_\mu +{\sqrt N' \over \sqrt N} b'_\mu
\ee
so that the term $\varepsilon^{\mu\nu\rho} b^-_\mu\partial_\nu b^-_\rho$ cancels. 
Since this $b^+_\mu$ does not couple to the matter fields, 
one can integrate it out by solving equation of motion. 
The equation of motion is 
\be
\varepsilon^{\mu\nu\rho}(2 \partial_\nu b^+_\rho +{2 (\sqrt{N'N} + N') \over N-N'}  \partial_\nu b^-_\rho)=0.
\ee
We can solve this by 
\be
b^+_\rho = - { (\sqrt{N'N} + N') \over N-N'} b^-_\rho.
\ee
Plugging back this into \eqref{csterm2} gives
\beal{
 \kappa (\cL_{cs}[A] - \cL_{cs}[A']) \to&  
  - i \bar\kappa \varepsilon^{\mu\nu\rho} b^-_\mu\partial_\nu b^-_\rho+\kappa (\cL_{cs}[A] - \cL_{cs}[A']), 
}
where $\bar\kappa$ is defined by
\be
\bar\kappa = \kappa { N N' \over N- N'}.
\ee
By collecting all the terms the whole action \eqref{actionbif} becomes 
\beal{
S\to& \int d^3 x \biggl( - i \bar\kappa \varepsilon^{\mu\nu\rho} b^-_\mu\partial_\nu b^-_\rho + \Tr[(b^-_\mu)^2 q^\dagger_A q^A + i b^-_\mu(q^\dagger_A D^\mu q^A - D^\mu q^\dagger _A q^A - \psi^\dagger{}^A \gamma^\mu \psi_A)]\nn
&+  \kappa (\cL_{cs}[A] - \cL_{cs}[A']) + \cL_{m} \biggl) .
\label{actionsplit}
}
In summary, the first line, which contains $b^-_\mu$, is coming from $U(1)\times U(1)$ part and the second one is $SU(N)\times SU(N)$ part. 

Now let us fix the gauge degrees of freedom by the light-cone gauge for $A_\mu, A_\mu',  b^-_\mu$
and integrate them out as done in the previous section. 
% For the contribution of $U(1)$ gauge field $b_\mu$, 
% we simply follow the prescription taken in \cite{Gaiotto:2007qi} by setting it to zero. 
While the equation of motion for $A_+$ has the same form as \eqref{eoma+}, 
those for the gauge fields $b^-_\mu, A'$ are 
\beal{
-2\bar\kappa \partial_- b^{-}_3 =&\Tr_{N'}\[ i\(q^\dagger _A \partial_- q^A - \partial_- q^\dagger_A q^A  
- \psi^\dagger{}^A \gamma_- \psi_A \)\],  \\
-2\kappa \partial_- A^{a'}_3 =&\Tr_{N'}\[ i\(q^\dagger _A \partial_- q^A - \partial_- q^\dagger_A q^A  
- \psi^\dagger{}^A \gamma_- \psi_A \) T^{a'} \], 
}
where $T^{a'}$ is a generator of $SU(N')$ gauge group and $\Tr_{N'}$ is trace for $N'\times N'$ matrix. 
In the Fourier space they become
\beal{
2\bar\kappa i q_- b^{-}_3(q) =&  {N} \int {d^3 r \over (2\pi)^3}\Tr_{N'}\[ \( (2r+q)_- \chi^A_A(q, r+ {q\over 2})
+ 2 i \xi_-{}^A_A(q, r+ {q\over 2})\) \], \nn
2\kappa i q_- A^{a'}_3(q) =&  {N} \int {d^3 r \over (2\pi)^3}\Tr_{N'}\[ \( (2r+q)_- \chi^A_A(q, r+ {q\over 2})
+ 2 i \xi_-{}^A_A(q, r+ {q\over 2})\) T^{a'}\],
\label{solutiona'3}
}
where we used the notation $\chi^A_B, \xi_-{}^A_B$  analogous to \eqref{chi}, \eqref{xi-}, which are now $N'$ by $N'$ matrices. 
Solving these and substituting back into \eqref{actionsplit} 
we find the analog of \eqref{s1}, which now also contains the contribution from the gauge fields $b^-, A'$. 

Then we introduce auxiliary fields to eliminate all the interactions by adding the terms, 
which has the same form as \eqref{deltaS} except the auxiliary fields are now $N' \times N'$ matrices 
and suitable contractions for $SU(N')$ indices. 
By this manipulation what we shall do is essentially to exchange $\chi, \xi$ into $\alpha, \beta$.  
For example the constraint equations of $b^-_3, A^{a'}_3$ become
\beal{
2\bar\kappa i q_- b^{-}_3(q) =&  {N} \int {d^3 r \over (2\pi)^3}\Tr_{N'}\[ \( (2r+q)_- \alpha^A_A(q, r+ {q\over 2})
+ 2 i \beta_-{}^A_A(q, r+ {q\over 2})\) \], \nn
2\kappa i q_- A^{a'}_3(q) =&  {N} \int {d^3 r \over (2\pi)^3}\Tr_{N'}\[ \( (2r+q)_- \alpha^A_A(q, r+ {q\over 2})
+ 2 i \beta_-{}^A_A(q, r+ {q\over 2})\) T^{a'}\].
\label{a'3}
}
After this treatment we can integrate out the elementary fields $q^A, \psi_A$ 
and obtain the analog of \eqref{s3}, which contains the contribution from $b^-, A'$.

Then we evaluate the action at saddle points to study the leading expression in the large $N$ limit. 
We assume the same ansatz for saddle pints such as translational, rotational invariance 
and covariance of flavor indices. 
In addition to these we also naturally expect saddle points to satisfy covariance of $SU(N')$ fundamental indices.   
\beal{
&\aver{\alpha^A_B(P,q)}=(2 \pi)^3 \delta^3(P) \delta^A_B {\bf 1}_{N'}\alpha(q), \;
\aver{\beta^A_B(P,q)}= (2 \pi)^3 \delta^3(P) \delta^A_B {\bf 1}_{N'}\beta(q),\; \nn
&\aver{\Sigma^A_B(P,q)}=(2 \pi)^3 \delta^3(P) \delta^A_B{\bf 1}_{N'}\Sigma(q), \;
\aver{\Pi^A_B(P,q)}= (2 \pi)^3 \delta^3(P) \delta^A_B{\bf 1}_{N'}\Pi(q),\; 
 \label{assumptions2}
 }
 where ${\bf 1}_{N'}$ is the $N' \times N'$ unit matrix. 
In other words,  
we treat $SU(N')$ gauge symmetry as flavor symmetry under the 't Hooft vector model limit. 
We set to zero for the fermionic fields. 
Under this ansatz 
consistent solutions for $b^-_3, A^{a'}_3$ in \eqref{a'3} become trivial. 
This is because under the ansatz \eqref{assumptions2} the right-hand side in \eqref{a'3} becomes 
proportional to $\delta^3(q)$ but the left-hand side $q_-$, which requires $b^-_3, A^{a'}_3$
to vanish. 
Note that $A^{a'}_3=0$ is also required from gauge index contraction of the right-hand side 
since $\Tr_{N'} T^{a'}=0$. 
Accordingly the right-hand side in \eqref{a'3} also has to vanish so that 
it is required to satisfy 
\be
\int {d^3 r \over (2\pi)^3} \( 2r_- \alpha(r) + 2 i \beta_-(r)\) =0.
\label{eomrhs}
\ee
This has to be checked after solving saddle point equations for $\alpha, \beta$, but 
we can check now because
we already know that the solutions of $\alpha, \beta$ are given by exact propagators of scalar and fermion 
respectively of the following form \cite{Jain:2012qi}
\be
\alpha(r) = {1 \over r^2 + c_{B,0}^2}, \quad \beta_-(r) = {i r_- \over r^2 + c_{F,0}^2}
\label{solab}
\ee 
where $c_{B,0}, c_{F,0}$ are pole masses of scalar and fermion respectively.
Plugging \eqref{solab} into \eqref{eomrhs} one can see that the left-hand side vanishes 
by performing the angular integral. 
As a result $U(N')$ sector does not contribute at all under the limit. 
In other words, 
$U(N')$ gauge factor is so weakly gauged as to decouple from the leading contribution  
under the 't Hooft vector model limit.
Thus the result of the effective action is essentially the same as that of the case with one gauge group \eqref{s4}. 
\beal{
S_{\text{eff}}=&NN'MV \biggl[ \int \frac{d^3q}{(2 \pi)^3} \left( \log (q^2 +\Sigma(q) )
-\tr\log ( i \gamma^\mu q_\mu + \Pi(q)) \right) \nn
&+ \int \frac{d^3q_1}{(2 \pi)^3} \frac{d^3q_2}{(2 \pi)^3} 
 C_1(q_1,q_2)  \alpha(q_1) \alpha(q_2) \nn
&+ \int \frac{d^3q_1}{(2 \pi)^3} \frac{d^3q_2}{(2 \pi)^3}\frac{d^3q_3}{(2 \pi)^3} C_2(q_1,q_2,q_3)
\alpha(q_1)\alpha(q_2)\alpha(q_3)\nn
&+\int \frac{d^3q_1}{(2 \pi)^3} \frac{d^3q_2}{(2 \pi)^3}~
\frac{8 \pi iN}{ k (q_1-q_2)_{-}} \beta{}_-(q_1) \beta{}_I(q_2)\nn
&- \int \frac{d^3 q}{(2 \pi)^3}
\biggl( \Sigma(q) \alpha(q) +  2\Pi{}^I(q) \beta{}_I(q) 
+ 2\Pi^-(q) \beta_-(q) \biggl) \biggl] \nn
&+ S_{m} (\alpha\delta^B_A, \beta{}_I\delta^B_A, \beta{}_-\delta^B_A,0,0).
\label{s5}
}
Let us apply this result to $\cN=4$ Chern-Simons-matter theory with $U(N)_k \times U(N')_{-k}$ case and 
$\cN=6$ with $U(N)_k \times U(N')_{-k}$ (ABJ) case.

%%%%%%%%%%%%%%%%%%%%%%%%%%%%%%%%%%%%%%%%%%%%%%%%%%%%%%
\subsubsection{$\cN=4$ $U(N)_k \times U(N')_{-k}$ case}

In this subsection we apply the result \eqref{s5} to $\cN=4$ $U(N)_k \times U(N')_{-k}$ 
Chern-Simons theory, whose matter content is two bi-fundamental complex scalar fields $q^A$ and fermionic fields $\psi_A$, 
where $A=1,2$.
This Chern-Simons-matter Lagrangian is given by \eqref{n4action}. 
The potential of the matter fields is 
\beal{
V^{\cN=4}_m=& \Tr \biggl[  {1 \over 2 \kappa}\left(q^\dagger_B q^B \psi^\dagger{}^{  A} \psi_{A}- q^Bq^\dagger_B \psi_{A}\psi^\dagger{}^{  A}
 -\varepsilon_{AC} \varepsilon_{BD} \psi^\dagger{}^A q^B \psi^\dagger{}^C q^D 
+\varepsilon^{AC} \varepsilon^{BD} \psi_A q^\dagger_B \psi_C q^\dagger_D \right) 
\nn
&+ {1 \over \kappa^2} 
\biggl({3\over 2}  q_A^\dagger  q^B q^\dagger_B q^A q^\dagger_C q^C
-{2\over 3} q^A  q^\dagger_B q^C q^\dagger_A q^B q^\dagger_C 
-{5\over 12} (q^A q_A^\dagger  q^B  q^\dagger_B  q^C  q^\dagger_C
+q_A^\dagger  q^A q^\dagger_B q^B q^\dagger_C q^C) \biggl) \biggl].
}
Thus $S_m$ in $\cN=4$ case is  
\beal{
&S^{\cN=4}_m (\chi^B_A, \xi^A_B{}_-, \xi^A_B{}_I, \eta_{AB}, \bar\eta^{AB})  \nn
=&N \int  \frac{d^3P}{(2 \pi)^3} \frac{d^3q_1}{(2 \pi)^3} \frac{d^3q_2}{(2 \pi)^3} 
 {N\over 2 \kappa} \Tr_{N'}\biggl[2 \xi^B_B{}_I(P,q_1) \chi^A_A(-P,q_2) 
- \eta_{BA}(P,q_1)  \bar\eta^{AB}(-P,q_2) \nn
& -  \varepsilon_{AB} \varepsilon_{CD}\bar\eta^{AB} (P,q_1)  \bar\eta^{CD} (-P,q_2) 
+  \varepsilon^{AB} \varepsilon^{CD} \eta_{AB} (P,q_1)  \eta_{CD} (-P,q_2) \biggl]
\nn
&+ N \int \frac{d^3P_1}{(2 \pi)^3} \frac{d^3P_2}{(2 \pi)^3}  \frac{d^3q_1}{(2 \pi)^3} \frac{d^3q_2}{(2 \pi)^3}\frac{d^3q_3}{(2 \pi)^3} {N^2 \over \kappa^2} \nn 
& \Tr_{N'}\biggl[ {3\over 2} \chi^B_A(P_1,q_1)\chi^A_B(P_2,q_2)\chi^C_C(-P_1-P_2,q_3)
 -{2\over 3} \chi^C_B(P_1,q_1)\chi^B_A(P_2,q_2)\chi^B_C(-P_1-P_2,q_3) \nn
&-{5\over 12} \( \chi^B_A(P_1,q_1)\chi^C_B(P_2,q_2)\chi^A_C(-P_1-P_2,q_3)  
+\chi^A_A(P_1,q_1)\chi^B_B(P_2,q_2)\chi^C_C(-P_1-P_2,q_3)\) \biggl],
}
Under the assumption \eqref{assumptions2} this reduces  
\beal{
&S^{\cN=4}_{m} (\alpha\delta^B_A, \beta{}_I\delta^B_A, \beta{}_-\delta^B_A,0,0) \nn
=&2NN'V \[ \int \frac{d^3q_1}{(2 \pi)^3} \frac{d^3q_2}{(2 \pi)^3} 
 {2N\over \kappa} \beta_I(q_1) \alpha(q_2) + \int \frac{d^3q_1}{(2 \pi)^3} \frac{d^3q_2}{(2 \pi)^3}\frac{d^3q_3}{(2 \pi)^3} 
 {N^2 \over (2\kappa)^2} \alpha(q_1)\alpha(q_2)\alpha(q_3) \],
}
which is $2N'$ times as that of the $\cN=2$ $U(N)_k$ Chern-Simons theory with minimal matter content:
\be
S^{\cN=4}_{m} (\alpha\delta^B_A, \beta{}_I\delta^B_A, \beta{}_-\delta^B_A,0,0)
=2N' S^{\cN=2}_{m} (\alpha, \beta{}_I, \beta{}_-,0,0).
\label{n2n4}
\ee

By taking \eqref{s5} into account, the total large $N$ effective action in the $\cN=4$ Chern-Simons theory with 
minimal matter content is exactly $2N'$ as that of the minimal $\cN=2$ Chern-Simons theory in the 't Hooft limit.

\subsubsection{Mass-deformed $\cN=4$ case} 

In this subsection we investigate large $N$ exact effective action of the previous $\cN=4$ 
Chern-Simons matter theory deforming the theory by a mass term keeping $\cN=4$ supersymmetry 
as well as $SO(4)$ R-symmetry \cite{Hosomichi:2008jd}. 
The $\cN=4$ mass term is given by \eqref{n4mass}
\beal{
\cL^{\cN=4}_{\text{mass}}
&= \Tr \biggl[  \mu \psi^\dagger{}^{  A}\psi_{A}
+ \mu^2   q_A^\dagger q^A + {\mu \over \kappa} ( (q^\dagger_A q^A)^2 - q^\dagger_A q^B q^\dagger_B q^A) \biggl]
}
where $\mu$ is a mass parameter. 
Since the mass term does not break the global symmetry, 
that of the vacuum is unchanged and thus the ansatz \eqref{assumptions2} holds. 
Under the ansatz this $\cN=4$ mass term becomes 
\beal{
S^{\cN=4}_{\text{mass}}   (\alpha, \beta{}_I, \beta{}_-)
&=2 NN' V \biggl [\int  \frac{d^3q_1}{(2 \pi)^3} ( 2\mu \beta_I(q_1) + \mu^2 \alpha(q_1))
+ \int\frac{d^3q_1}{(2 \pi)^3} \frac{d^3q_2}{(2 \pi)^3} { \mu \over \kappa} \alpha(q_1) \alpha(q_2) \biggl],
}
which is completely the same as the term obtained from the $\cN=2$ mass term 
\eqref{n1mass} with $w=1$ reduced 
under the assumption \eqref{assumptions2} with the over all multiplicative factor $2N'$:
\be
S^{\cN=4}_{\text{mass}}   (\alpha, \beta{}_I, \beta{}_-) = 2N' S^{\cN=2}_{\text{mass}}  (\alpha, \beta{}_I, \beta{}_-).
\ee

As a result the relation of the exact effective actions for $\cN=2$ and $\cN=4$ given by \eqref{n2n4} 
is unchanged under the $\cN=2$ and $\cN=4$ mass deformations. 
Again, $\cN=4$ effective action reduces to that of $\cN=2$ with an appropriate factor including 
the mass terms keeping the same amount of supersymmetry.

%%%%%%%%%%%%%%%%%%%%%%%%%%%%%%%%%%%%%%%

\subsubsection{$\cN=6$ $U(N)_k \times U(N')_{-k}$ (ABJ) case}

In this subsection we apply the result \eqref{s5} to ABJ theory, 
whose matter content is four bi-fundamental complex scalar fields $Y^A$ and fermionic fields $\Psi_A$, 
where $A=1,2,3,4$. 
This theory possesses $SU(4)$ R-symmetry and $U(1)_b$ global symmetry. 
The Lagrangian of ABJ theory is given by \eqref{n6action}. 
The potential of the matter fields is 
\beal{
V^{\cN=6}_m=& \Tr \biggl[ {1 \over 2 \kappa } ( Y^\dagger_A Y^A \Psi^\dagger{} ^B \Psi_B - Y^A Y^\dagger_A \Psi _B
\Psi^\dagger{} ^B+2 Y^A Y^\dagger_B \Psi _A \Psi^\dagger{} ^B - 2 Y^\dagger _A Y^B \Psi^\dagger{} ^A
\Psi_B \nn
&\quad -\varepsilon_{ABCD} \Psi^\dagger{}^A Y^B \Psi^\dagger{}^C Y^D 
+\varepsilon^{ABCD} \Psi_A Y^\dagger_B \Psi_C Y^\dagger_D ) \nn
&+{1 \over 12 \kappa^2 } \biggl( 
- Y^\dagger_A Y^A Y^\dagger_B Y^B Y^\dagger_C Y^C - Y^A Y^\dagger_A Y^B Y^\dagger_B Y^C Y^\dagger_C
- 4 Y^A Y^\dagger_B  Y^C  Y^\dagger_A Y^B Y^\dagger_C \nn
&\quad +6  Y^A Y^\dagger_B Y^B Y^\dagger_A Y^C Y^\dagger_C\biggl) \biggl].
}
$S_m$ in $\cN=6$ case is  
\beal{
&S^{\cN=6}_m (\chi^B_A, \xi^A_B{}_-, \xi^A_B{}_I, \eta_{AB}, \bar\eta^{AB})  \nn
=&N \int  \frac{d^3P}{(2 \pi)^3} \frac{d^3q_1}{(2 \pi)^3} \frac{d^3q_2}{(2 \pi)^3} 
 {N\over 2 \kappa} \Tr_{N'}\biggl[
 2 \chi^A_A(P,q_1) \xi^B_B{}_I(-P,q_2) - \eta_{AB}(P,q_1)  \bar\eta^{BA}(-P,q_2) \nn
&+ 2\eta_{BA}(P,q_1)  \bar\eta^{BA}(-P,q_2) - 4\chi^B_A{}(P,q_1) \xi^A_B{}_I(-P,q_2) \nn
& -  \varepsilon_{ABCD}\bar\eta^{AB} (P,q_1)  \bar\eta^{CD} (-P,q_2) 
+  \varepsilon^{ABCD} \eta_{BC} (P,q_1)  \eta_{DA} (-P,q_2) \biggl]
\nn
&+ N \int \frac{d^3P_1}{(2 \pi)^3} \frac{d^3P_2}{(2 \pi)^3}  \frac{d^3q_1}{(2 \pi)^3} \frac{d^3q_2}{(2 \pi)^3}\frac{d^3q_3}{(2 \pi)^3} {N^2 \over 12 \kappa^2} \nn 
& \Tr_{N'}\biggl[ - \chi^A_A(P_1,q_1)\chi^B_B(P_2,q_2)\chi^C_C(-P_1-P_2,q_3)
 - \chi^B_A(P_1,q_1)\chi^C_B(P_2,q_2)\chi^A_C(-P_1-P_2,q_3) \nn
&-4 \chi^C_B(P_1,q_1)\chi^B_A(P_2,q_2)\chi^A_C(-P_1-P_2,q_3)  
+6 \chi^B_B(P_1,q_1)\chi^C_A(P_2,q_2)\chi^A_C(-P_1-P_2,q_3) \biggl].
\label{smn6}
}
Under the assumption \eqref{assumptions2} this reduces  
\beal{
&S^{\cN=6}_{m} (\alpha\delta^B_A, \beta{}_I\delta^B_A, \beta{}_-\delta^B_A,0,0) \nn
=&4NN'V \[ \int \frac{d^3q_1}{(2 \pi)^3} \frac{d^3q_2}{(2 \pi)^3} 
 {2N\over \kappa} \beta_I(q_1) \alpha(q_2) + \int \frac{d^3q_1}{(2 \pi)^3} \frac{d^3q_2}{(2 \pi)^3}\frac{d^3q_3}{(2 \pi)^3} 
 {N^2 \over (2\kappa)^2} \alpha(q_1)\alpha(q_2)\alpha(q_3) \]
}
which is $4N'$ times as that of the $\cN=2$ $U(N)_k$ Chern-Simons theory with minimal matter content:
\be
S^{\cN=6}_{m} (\alpha\delta^B_A, \beta{}_I\delta^B_A, \beta{}_-\delta^B_A,0,0)
=4N' S^{\cN=2}_{m} (\alpha, \beta{}_I, \beta{}_-,0,0).
\label{n3n6}
\ee

By taking \eqref{s5} into account, the total large $N$ effective action in ABJ theory 
is exactly $4N'$ times that of the minimal $\cN=2$ Chern-Simons theory in the 't Hooft limit. 

The reason why the large $N$ effective action of ABJ theory has reduced to that of the $\cN=2$ one 
will be the same as in the $\cN=3$ case discussed in Section \ref{n3}. 
The ABJ(M) action can be constructed by using $\cN=2$ superfield formulation \cite{Benna:2008zy}. 
In the notation of \cite{Benna:2008zy}, 
the superpotential of the $\cN=6$ theory is of the form 
$
W^{\cN=6} \sim \varepsilon_{AC}\varepsilon^{BD}\Tr (Z^A W_{B} Z^C W_{D} )
$
up to some over all factor, 
where $Z^A, W_B\, (A,B=1,2)$ are bi-fundamental, anti-bi-fundamental chiral superfields, respectively.
Under the 't Hooft vector model limit, the contribution of the superpotential to effective action is given by 
$
\aver{W^{\cN=6}} \sim \varepsilon_{AC}\varepsilon^{BD}
\Tr (\aver{Z^A W_{B}} \aver{Z^C W_{D}} )
$
due to the large $N$ factorization, where $Z^A W_{B}$ is an $N'\times N'$ matrix.
However this contribution vanishes 
under the $SU(4)$ symmetric vacuum configuration \eqref{assumptions2}, 
which will explain the reduction observed above.

\subsubsection{Mass-deformed ABJ case} 

In this subsection we study large $N$ exact effective action in ABJ model deformed 
by a mass term keeping $\cN=6$ supersymmetry \cite{Gomis:2008vc}. 
The $\cN=6$ mass term is given by \eqref{n6mass}
\beal{
\cL^{\cN=6}_{\text{mass}}
&= \Tr \biggl[  \mu \psi^\dagger{}^{  A} M^B_A\psi_{  B}
+ \mu^2   Y_A^\dagger  M^A_B Y^B + {\mu \over \kappa} ( Y^\dagger_B M_A^B Y^A Y^\dagger_C Y^C -Y^A M_A^B Y^\dagger_B Y^C Y^\dagger_C) \biggl]
}
where $\mu$ is a mass parameter and $M_{A}^{B} = \text{diag}(1,1,-1,-1)$. 
This mass term breaks the $SU(4) \times U(1)_b$ global symmetry to 
$SU(2) \times SU(2) \times U(1) \times \bZ _2$ and thus changes the vacuum structure.  
A plausible ansatz respecting the global symmetry may be the following. 
\beal{
\aver{\alpha^A_B(P,q)}
=&(2 \pi)^3 \delta^3(P) {\bf 1}_{N'}(\delta^A_B \alpha_1(q) + M^A_B\alpha_2(q)) , \\
\aver{\beta^A_B(P,q)}=& (2 \pi)^3 \delta^3(P) {\bf 1}_{N'}(\delta^A_B\beta_1(q)+M^A_B\beta_2(q)),\\
\aver{\Sigma^A_B(P,q)}=&(2 \pi)^3 \delta^3(P){\bf 1}_{N'}(\delta^A_B\Sigma_1(q)+M^A_B\Sigma_2(q)),\\
\aver{\Pi^A_B(P,q)}=& (2 \pi)^3 \delta^3(P){\bf 1}_{N'}(\delta^A_B\Pi_1(q)+M^A_B\Pi_2(q)),\; 
 \label{assumptions3}
 }
 where $\alpha_i, \beta_i, \Sigma_i, \Pi_i \, (i=1,2)$ are determined by saddle point equations. 
Under this ansatz the effective action corresponding to \eqref{s5} becomes 
\beal{
S^{\cN=6}_{\text{eff}}=&NN'V \biggl[ \int \frac{d^3q}{(2 \pi)^3} 2 \biggl( \log (q^2 +\Sigma_1(q)+\Sigma_2(q) )
+ \log (q^2 +\Sigma_1(q)-\Sigma_2(q) ) \nn
&\qquad -\tr\log ( i \gamma^\mu q_\mu +  \Pi_1(q)+\Pi_2(q))  
-\tr\log ( i \gamma^\mu q_\mu +  \Pi_1(q)-\Pi_2(q)) \biggl) \nn
&+ \int \frac{d^3q_1}{(2 \pi)^3} \frac{d^3q_2}{(2 \pi)^3} 
 4 C_1(q_1,q_2)  \biggl(\alpha_1(q_1) \alpha_1(q_2)+\alpha_2(q_1) \alpha_2(q_2)\biggl) \nn
&+ \int \frac{d^3q_1}{(2 \pi)^3} \frac{d^3q_2}{(2 \pi)^3}\frac{d^3q_3}{(2 \pi)^3} 
4 C_2(q_1,q_2,q_3) \biggl( \alpha_1(q_1)\alpha_1(q_2)\alpha_1(q_3) +3 \alpha_1(q_1)\alpha_2(q_2)\alpha_2(q_3)  \biggl)\nn
&+\int \frac{d^3q_1}{(2 \pi)^3} \frac{d^3q_2}{(2 \pi)^3}~4\times 
\frac{8 \pi iN}{ k (q_1-q_2)_{-}}\biggl( \beta_1{}_-(q_1) \beta_1{}_I(q_2)+ \beta_2{}_-(q_1) \beta_2{}_I(q_2)\biggl) \nn
&- \int \frac{d^3 q}{(2 \pi)^3} 4
\biggl(  \Sigma_1(q) \alpha_1(q) + \Sigma_2(q) \alpha_2(q) +  2\Pi_1{}^I(q) \beta_1{}_I(q) +  2\Pi_2{}^I(q) \beta_2{}_I(q) \nn 
&\qquad + 2\Pi_1^-(q) \beta_1{}_-(q) + 2\Pi_2^-(q) \beta_2{}_-(q) \biggl) \biggl]+ S_{m}^{\cN=6}
\label{s6}
}
where $S_m$ is 
\beal{
S_m^{\cN=6}
% =&NM' V \biggl [\int  \frac{d^3q_1}{(2 \pi)^3} (4 \mu \beta_2{}_I(q_1) + 4 \mu^2 \alpha_2(q_1))
% + \int\frac{d^3q_1}{(2 \pi)^3} \frac{d^3q_2}{(2 \pi)^3}8 { \mu \over \kappa} \alpha_1(q_1) \alpha_2(q_2)  \nn
% &+\int  \frac{d^3q_1}{(2 \pi)^3} \frac{d^3q_2}{(2 \pi)^3} 
%  {N\over 2 \kappa} \biggl(
%  2\cdot 4^2 \alpha_1(q_1) \beta_1{}_I(q_2) 
%  - 4\cdot 4( \alpha_1(q_1) \beta_1{}_I(q_2) +\alpha_2(q_1)\beta_2{}_I(q_2)) \biggl)
% \nn
% &+\int  \frac{d^3q_1}{(2 \pi)^3} \frac{d^3q_2}{(2 \pi)^3}\frac{d^3q_3}{(2 \pi)^3} {N^2 \over 12 \kappa^2} \nn 
% & \biggl( - 4^3 \alpha_1(q_1)\alpha_1(q_2)\alpha_1(q_3)
%  -5 \cdot 4 ( \alpha_1(q_1) \alpha_1(q_2)\alpha_1(q_3) +3 \alpha_1(q_1) \alpha_2(q_2)\alpha_2(q_3) )\nn
% &\qquad +6\cdot 4^2\alpha_1(q_1) ( \alpha_1(q_2)\alpha_1(q_3) + \alpha_2(q_2)\alpha_2(q_3) ) 
% \biggl) \biggl] \nn
=&NM' V \biggl [\int  \frac{d^3q_1}{(2 \pi)^3} 4( 2\mu \beta_2{}_I(q_1) + \mu^2 \alpha_2(q_1))
+ \int\frac{d^3q_1}{(2 \pi)^3} \frac{d^3q_2}{(2 \pi)^3}8 { \mu \over \kappa} \alpha_1(q_1) \alpha_2(q_2)  \nn
&+\int  \frac{d^3q_1}{(2 \pi)^3} \frac{d^3q_2}{(2 \pi)^3} 
 4^2 \times {N\over 2 \kappa} \biggl(
  \alpha_1(q_1) \beta_1{}_I(q_2) 
 - \alpha_2(q_1)\beta_2{}_I(q_2)  \biggl)
\nn
&+\int  \frac{d^3q_1}{(2 \pi)^3} \frac{d^3q_2}{(2 \pi)^3}\frac{d^3q_3}{(2 \pi)^3} {N^2 \over 12 \kappa^2}  \biggl( 12 \alpha_1(q_1)\alpha_1(q_2)\alpha_1(q_3) +36 \alpha_1(q_1) \alpha_2(q_2)\alpha_2(q_3) 
\biggl) \biggl], 
}
which includes the mass term.
Let us rewrite this in terms of the following variables
\beal{
\Sigma^{(\pm)} = \Sigma_1 \pm \Sigma_2, \quad 
\Pi^{(\pm)} = \Pi_1 \pm \Pi_2, \quad 
\alpha^{(\pm)} = \alpha_1 \pm \alpha_2, \quad 
\beta^{(\pm)} = \beta_1 \pm \beta_2, \quad 
}
which can simplify \eqref{s6}. 
A simple form of the large $N$ effective action for massive ABJ case is the following.  
\beal{
S^{\cN=6}_{\text{eff}} =&2N' \biggl( S^{\cN=2}_{\text{eff}}(\alpha^{(+)}, \beta^{(+)}_I, \beta^{(+)}_-,0,0)
|_{\Sigma\to\Sigma^{(+)}, \Pi\to\Pi^{(+)}}
+ S^{\cN=2}_{\text{eff}}(\alpha^{(-)}, \beta^{(-)}_I, \beta^{(-)}_-,0,0) |_{\Sigma\to\Sigma^{(-)}, \Pi\to\Pi^{(-)}}
 \nn
&+ S^{\cN=2}_{\text{mass}}(\alpha^{(+)}, \beta^{(+)}_I, \beta^{(+)}_-)
-S^{\cN=2}_{\text{mass}}(\alpha^{(-)}, \beta^{(-)}_I, \beta^{(-)}_-) \biggl)\nn
&-NV\int  \frac{d^3q_1}{(2 \pi)^3} \frac{d^3q_2}{(2 \pi)^3} 
 {2 N\over  \kappa} ( \alpha^{(+)}(q_1) - \alpha^{(-)}(q_2) ) ( \beta_I^{(+)}(q_1) - \beta_I^{(-)}(q_2) ). 
 \label{seffn6massive}
 }
 
 We observe a splitting of the mass term in the effective action due to 
 the fact that the $\cN=6$ mass term breaks $SU(4)$ R-symmetry to two $SU(2)$s.  
 Determining the saddle point equations and solving them is beyond the scope of this paper. 
As a trivial check we can see that in the case with $\mu=0$ this effective action reduces to that of massless ABJ, 
% which is $4N'$ multiple of effective action of $\cN=2$ minimal Chern-Simons vector model, 
because we have a solution $ \alpha^{(+)}= \alpha^{(-)},  \beta^{(+)}= \beta^{(-)}$. 
 
% Note that 
% \beal{
%  \alpha_1\alpha_1\alpha_1 +3 \alpha_1\alpha_2\alpha_2
%  =& \half (\alpha^{(+)}\alpha^{(+)}\alpha^{(+)} +\alpha^{(-)}\alpha^{(-)}\alpha^{(-)} ) \nn 
%  } 

%%%%%%%%%%%%%%%%%%%%%%%%%%%%%%%%%%%%%%%
\section{Comments on thermal free energy and duality} 
\label{comments}

In the previous section 
we have obtained the large $N$ exact effective actions for $\cN=3,4,6$ Chern-Simons 
matter theories.
Once one obtains large $N$ exact effective actions 
one can compute exact large $N$ thermal free energies at an arbitrary temperature  
by performing Wick rotation for time direction and compactifying the Euclidean time in a circle 
whose circumference is the inverse temperature.
Due to appearance of circle 
one has to care about boundary conditions and holonomy. 
We set boundary conditions for this circle such that the scalar fields satisfy periodic one 
and the fermionic fields do anti-periodic one to study thermal canonical ensemble of the system. 
According to the boundary conditions, we exchange the integration of the momentum for the time direction 
into the summation over the discrete Fourier modes satisfying suitable boundary conditions. 
The holonomy is zero mode of gauge field on the circle and
it can be taken into account by implementing a constant shift by holonomy 
for the thermal-time component of momentum appearing in the propagators \cite{Yokoyama:2012fa}. 
We normalize a thermal free energy in such a way that 
it vanishes at zero temperature. 

For holonomy configuration determined by minimizing the free energy, 
a crucial argument was made in \cite{Aharony:2012ns}
that each eigenvalue of holonomy matrix obeys the fermionic statistics
in the high temperature limit 
so that the holonomy configuration does not cramp but spread 
around the origin with the width ${2\pi\lambda}$ and height ${1 \over 2\pi \lambda}$
in the 't Hooft large $N$ limit for $U(N)$ level $k$ Chern-Simons theory with one fundamental 
boson, fermion or both. 
This can be confirmed not only from the canonical formalism 
but also from the path integral formalism \cite{Jain:2013py}. 
Taking account of this holonomy effect 
one can see three dimensional duality of this class of the theories at a high temperature
of order $\sqrt N$. 

Let us consider the holonomy distribution in the situations of this paper. 
First let us consider $U(N)$ level $k$ Chern-Simons theory with any finite number of fundamental 
fields. 
Under the 't Hooft limit holding the number of matter fields fixed 
the holonomy distribution clearly becomes the same as that in one fundamental flavor case. 
This implies, from the calculation in the previous section, 
the large $N$ free energy of $U(N)_k$ $\cN=3$ Chern-Simons theory with one pair of fundamental and 
anti-fundamental chiral fields (quark and anti-quark) precisely reduces to twice of that of $U(N)_k$
$\cN=2$ Chern-Simons theory with one chiral fundamental multiplet. 
Since this $\cN=2$ Chern-Simons theory is self-dual under the exchange of
$\lambda$ and $\lambda-\text{sgn}(\lambda)$ \cite{Benini:2011mf,Aharony:2012ns}, 
this result suggests 
the minimal $\cN=3$ Chern-Simons theory is also self-dual under the same transformation of $\lambda$.

One may discuss this self-duality of $\cN=3$ in the following way.
For this purpose we first consider $\cN=2$ $U(N)_k$ Chern-Simons theory with $N_F$ quark flavors 
$(Q^i, \wt Q_j)$ with no superpotential. We call this electric theory for convenience. 
The dual of this theory, which we will call magnetic theory, 
is known as $\cN=2$ $U(N_F+|k|-N)_k$ Chern-Simons theory 
with $N_F$ dual quark flavors denoted by $(q^i, \wt q_j)$ 
and gauge-singlet fields $M_i^j$ with superpotential $\wt W_0 = \wt q_j M^j_i q^i$. 
These two theories are considered to be equivalent in the infra-red fixed point \cite{Giveon:2008zn}.
Now consider the case with $N_F=1$. 
Let us add a (marginally) relevant double trace chiral term in the superpotential  
$\Delta W= (\wt Q Q)^2$ in the electric theory and flow it to the
$\cN=3$ Chern-Simons theory \cite{Gaiotto:2007qi} as discussed in Section \ref{n3}.
What is the corresponding deformation in the magnetic side? 
The answer is to add the superpotential of the form $\Delta\wt W =  M^2$, 
since $M$ corresponds to the mesonic field in the electric side \cite{Giveon:2008zn}. 
Clearly this gives the mass term for the field $M$, which decouples in the IR. 
Integrating $M$ out gives a double trace chiral term in the superpotential of the magnetic theory. 
Therefore the resulting IR theory of the magnetic side also achieves $\cN=3$ supersymmetry
by using the argument of \cite{Giveon:2008zn}, 
which will account for the self-duality of the minimal $\cN=3$ theory.

Next we consider the holonomy distribution for $U(N)_k \times U(N')_{-k}$ Chern-Simons theory with any finite number of (bi-)fundamental fields. 
One has to take care of holonomy not only for $U(N)$ but also $U(N')$ in general $N$. 
But under the 't Hooft large $N$ limit keeping $N'$ and number of (bi-)fundamental fields fixed 
the contribution of holonomy for $U(N')_{-k}$ reduces to trivial one  
and that for $U(N)_k$ becomes the same as that for Chern-Simons theory with one fundamental 
flavor in the leading of large $N$ limit. 
Therefore, as happened in $\cN=3$ case, 
the free energy of $\cN=4$ Chern-Simons theory (including the $\cN=4$ mass term) and 
that of ABJ theory reduce to those of $\cN=2$ Chern-Simons theory with one chiral multiplet 
(including the $\cN=2$ mass term) up to overall integral factor. 
This suggests self-duality of $\cN=4$ theory including the $\cN=4$ mass term and ABJ theory. 
The self-duality of ABJ theory was already discussed in the original paper \cite{Aharony:2008gk}.
Their claim is $\cN=6$ theories with gauge group $U(N)_k \times U(N')_{-k}$ and 
$U(N')_k\times U(2N' + |k| -N)_{-k}$ are equivalent. 
Under the 't Hooft large $N$ limit with other parameters fixed 
this claim tells us that the physical quantities become the same under exchange of $\lambda$ 
with $ \lambda - \text{sgn}(\lambda)$, 
which is the same self-duality transformation as that of $\cN=2$ case. 
Our result gives strong evidence for this conjecture in a non-supersymmetric situation 
by confirming match of the large $N$ thermal free energy under the duality transformation.%

One may presumably perform analogous discussion 
of the self-duality of $\cN=4$ Chern-Simons theory including the case of finite $N$, 
but we leave further discussion in future.

The same holonomy distribution is also the case to mass-deformed ABJ theory under the limit. 
From the calculation in the previous section 
we observed the large $N$ effective action does not precisely reduce to 
that of $\cN=2$ with one chiral field, so neither does the large $N$ thermal free energy. 
Therefore it is not obvious to see self-duality of ABJ model with $\cN=6$ mass term
from our calculation. 
It is intriguing to explore this more by using not only the large $N$ thermal free energy 
but also other tools such as three-sphere partition function. 
We leave detailed analysis to future work.

%%%%%%%%%%%%%%%%%%%%%%%%%%%%%%%%%%%%%%%
\section{Discussion}
\label{discussion}

In this paper we have computed the effective actions and thermal free energies for $\cN=3$
$U(N)_k$ and $\cN=4,6$ $U(N)_k\times U(N')_{-k}$ 
Chern-Simons theories with minimal matter content including the $\cN=4,6$ mass term 
exactly in the 't Hooft large $N$ limit with the other parameters fixed. 
Under this limit all of them have reduced to the effective action 
or thermal free energy for $\cN=2$ with one chiral 
multiplet with the overall factor $M N'$, where $M$ is the number of a chiral or anti-chiral field
($N'=1$ for $\cN=3$ case), 
except the mass-deformed ABJ case. 
We have demonstrated that the self-duality of $\cN=3,4,6$ Chern-Simons theories 
(including the $\cN=4$ mass term) 
reduces to that of $\cN=2$ with one chiral field (including the $\cN=2$ mass term). 

In Section \ref{bifund} 
we have shown that there is no leading contribution of the $U(N')$ gauge fields under the 't Hooft 
vector model limit.  
As a result 
we observed that the resulting thermal free energy showed expected duality in Chern-Simons matter 
theories in the limit. 
This result also supports the prescription given in \cite{Gaiotto:2007qi} 
to deal with gauge fields in the large number of flavor limit in study of 
thermal free energy in Chern-Simons matter theories. 
However, at a finite Chern-Simons level $k$ there will be non-trivial contribution of $U(N')$ gauge fields. 
Especially the contribution of $U(1)$ part of the gauge fields will be important 
to see the relation between a Chern-Simons matter theory and the dual M-theory 
because the dual scalar field obtained by dualizing the $U(1)$ gauge field 
represents M-circle of the dual M-theory with the radius of order $1/k$. 
% But in the large $k$ limit this contribution is presumably negligible.   

There is a straight-forward generalization of the results of this paper by including chemical potential 
as done in \cite{Yokoyama:2012fa,Jain:2013gza}. 
Under the duality transformation 
chemical potential for scalar fields exchanges with that for fermionic fields.
But physics by including chemical potential is not so simple because 
it possibly gives rise to condensation of bosonic fields known as Bose-Einstein condensation
and Fermi surface of fermionic fields, 
which is perhaps unstable by something like the Cooper instability \cite{Cooper:1956zz}. 
It was observed that 
the duality works in the region where both bosonic and fermionic theories 
are in the uncondensed phase but the duality becomes unclear in the condensed phase \cite{Jain:2013gza}. 
It is interesting to explore the duality structure beyond the uncondensed phase. 

A technical but important issue is to calculate the next sub-leading correction of these theories
by concurrently taking large $M$ or large $N'$ limit keeping $M/N$ or $N'/N$ fixed. 
(Some perturbative calculation was done in \cite{Banerjee:2013nca}.)
Especially the large $N'$ limit is worthwhile to study properties
beyond the vector model limit of this class of Chern-Simons matter theories. 
Under the large $N'$ limit 
one has to take care of not only non-planar diagrams but also the sub-leading correction of holonomy 
distribution for $U(N')$ as well as $U(N)$. 
It is quite non-trivial to check whether the three dimensional duality holds up to the next leading 
of large $N$ limit. 
Since the Chern-Simons system reduces to $U(N')$ matrix model under the limit, 
the Vandermonde measure factor will play an important role 
to determine the correct holonomy distribution.%
\footnote{The author thanks S. P. Wadia for pointing this out. }
It is interesting to study how the $1/N$ corrected holonomy distribution and 
the behavior thereof under the duality transformation is modified from that found in \cite{Jain:2013py}. 

It is of interest to explore mass-deformed Chern-Simons vector models 
as in \cite{Jain:2013gza,Frishman:2013dvg}. 
In particular, a mass-deformed theory is free from infra-red divergence   
so that one can safely consider a scattering matrix.
It is interesting to determine S-matrix of elementary particles 
perturbatively and exactly in the 't Hooft large $N$ limit 
as done in correlation functions of conserved currents 
\cite{Giombi:2011rz,Maldacena:2011jn,Maldacena:2012sf,Aharony:2012nh,GurAri:2012is}. 
(See \cite{Nizami:2013tpa} for a recent computation of supersymmetric correlation functions.)

It is also interesting to study a gravity theory dual to Chern-Simons vector models in the context of 
$AdS_4/CFT_3$ correspondence, 
which is conjectured as a parity violating Vasiliev theory \cite{Vasiliev:1990en} 
on $AdS_4$ background 
with suitable boundary conditions \cite{Giombi:2011kc,Chang:2012kt}. 
(The original proposal was done in \cite{Klebanov:2002ja}. 
Related studies are, for example,  \cite{Sezgin:2002rt,Sezgin:2003pt,Giombi:2009wh,Giombi:2010vg,Giombi:2011ya}.
See also \cite{Giombi:2012ms,Sezgin:2012ag,Giombi:2013yva,Giombi:2013fka,Tseytlin:2013jya}
for reviews and recent computations of higher spin theories.)
According to \cite{Chang:2012kt}, 
a higher spin gravity theory dual to an $U(N)_k\times U(N')_{-k}$ 
bi-fundamental Chern-Simons theories such as ABJ theory 
is constructed from higher spin fields with $U(N')$ gauge indices. 
Therefore one can expand the bulk theory by a new bulk 't Hooft coupling 
$N'/N$ by taking the large $N', N$ limit with their ratio fixed.   
This indicates new confinement/deconfinement transition for higher spin fields 
with respect to the $U(N')$ gauge interaction. 
The field theory analysis by using a toy model in \cite{Chang:2012kt} suggested that 
the $U(N')$ gauge deconfinement happens at temperature of order one 
while the Hawking-page one occurs at temperature of order $\sqrt{N/ N'}$. 
This implies that as $N'$ goes to $N$, the higher spin fields become heavier so that
the $U(N')$ confinement/deconfinement phase transition point and the 
Hawaking-Page one for higher spin fields coalesce into 
the Hawking-Page one in non-higher spin gravity theory on a certain $AdS_4$ background.%
\footnote{For example, the non-higher spin gravity theory dual to ABJ theory is 
 type IIA supergravity theory on $AdS_4\times CP^3$ with a certain B-field background.} 
% To see these phenomena it will be important to investigate the $1/N$ or $N'/N$ correction. 
To realize the bulk picture proposed in \cite{Chang:2012kt},
it is important to clarify the confining mechanism of $U(N')$ gauge symmetry 
in the bulk, which may be different from that in usual QCD. 
This is simply because the bulk 't Hooft coupling $N'/N$ will not get renormalized.
As a result the dimensional transmutation which is often expected to happen in 
four dimensional Yang-Mills theories may not happen here.  
It is of interest to study how to compute the dynamical scale 
in the bulk $U(N')$ gauge theory and obtain the phase diagram thereof. 

We hope this note will become useful to address these issues in the future. 

%%%%%%%%%%%%%%%%%%%%%%%%%%%%%%%%%%%%%%%
\section*{Acknowledgments} 

The author would like to thank Spenta R. Wadia and especially Shiraz Minwalla 
for many useful discussions. 
He is also grateful to Sandip P. Trivedi, Spenta R. Wadia 
for reading the draft and giving helpful comments.

%%%%%%%%%%%%%%%%%%%%%%%%%%%%%%%%%%%%%%%%%%%%%%%
\appendix 
\section{Supersymmetric Chern-Simons-matter action}

In this section we present $\cN=1,2,3,4,6$ supersymmetric Chern-Simons-matter action of the minimal matter content with the mass term preserving the same amount of supersymmetry in our convention.

\subsection{$\cN=1$} 
\label{n1}

$\cN=1$ $U(N)_k$ Chern-Simons-matter action with one chiral multiplet $(q, \psi)$ in the fundamental representation of the gauge group is given in \cite{Jain:2012qi}.
The action is given by
\beal{
S^{{\cal N}=1}  
&= \int d^3 x \biggl[ i\kappa\varepsilon^{\mu\nu\rho}\Tr ( A_\mu\partial_\nu A_\rho -{2i\over3} A_\mu A_\nu A_\rho)  
 + D_\mu \bar  q D^\mu  q + \bar\psi \slash\!\!\!\! D \psi \nn
&\qquad + (\bar  q  q) W'^2_{\bar  q  q}+ (\bar\psi \psi) \left( - W'_{\bar  q  q}+{1  \over 2\kappa} (\bar  q  q)\right) -(\bar  q \psi) (\bar \psi  q) W''_{\bar  q q} \nn
&\qquad  
+\left(- \half{W''_{\bar  q  q}} - {1 \over 4 \kappa}\right) ((\bar\psi  q)( \bar \psi  q ) +(\bar  q \psi)( \bar  q \psi )) \biggl],
}
where $\sla D=\gamma^\mu D_\mu$,  $W_{\bar q  q}$ is a superpotential and $W'_x = {d W_x \over d x}$.
$\kappa$ is related to the Chern-Simons level $k$ by $\kappa =  {k \over 4\pi}$. 
The covariant derivative acts as \eqref{fcovderiv}. 
The contraction of gauge indices is understood by our bracket notation. For example,  
$(\bar  q  \psi)= \bar  q_m  \psi^m$, where $m$ is the fundamental gauge index. 
%and the covariant derivative acts on the fields as follows. 
% \beal{
% D_\mu q^A& = \partial_\mu q^A -i A_\mu q^A, \quad
% D_\mu q^\dagger_A= \partial_\mu q^\dagger_A +i q^\dagger_A  A_\mu,\\
% D_\mu \psi_A&= \partial_\mu\psi_A -i A_\mu \psi_A,\quad
% D_\mu \psi^\dagger{}^{A}= \partial_\mu\psi^\dagger{}^{A}+i \psi^\dagger{}^{A} A_\mu.
% \label{covderiv2}
% }  
Supersymmetry transformation rule is
\bea
\delta_\epsilon  q&=& -\sqrt 2 \epsilon\psi,\quad 
\delta_\epsilon \bar  q= -\sqrt 2 \epsilon\bar \psi,\\
\delta_\epsilon \psi_\alpha&=& 
\sqrt 2 (\epsilon_\beta \sla D^{\beta}{}_{\alpha} q -\epsilon_\alpha  q W'(\bar  q  q)),\\
\delta_\epsilon \bar\psi_\alpha&=& 
\sqrt 2 (\epsilon_\beta \sla D^{\beta}{}_{\alpha}\bar  q-\epsilon_\alpha W'(\bar  q  q) \bar  q),\\
\delta _\epsilon A_{\mu}&=&
{i \over \sqrt 2 \kappa} ( \epsilon \gamma_\mu \bar\psi q - \bar q \epsilon \gamma_\mu  \psi ).
 \eea
Hereafter we shall suppress the spinor indices $\alpha, \beta$. 

Superconformal action can be obtained by restricting the superpotential to be quadratic.
\be
W(\bar q q) = - {w \over 4 \kappa} (\bar q  q)^2
\label{n1superpot}
\ee
where $w$ is a real number. By putting $W'_{\bar q q} = - {w \over 2 \kappa} \bar q  q, 
W''_{\bar q q} = - {w \over 2 \kappa}$ above, we obtain the superconformal $\cN=1$ action and supersymmetry transformation.

On the other hand, 
the $\cN=1$ mass term can be obtained by adding a linear term in the superpotential. 
\be
W_{\text{mass}}(\bar  q  q) = -\mu \bar  q  q,
\label{superpotmass}
\ee
which is of the following form in the action:
\bea
S^{\cN=1}_{\text{mass}}
&=& \int d^3 x  [\mu^2 \bar q q + \mu \bar\psi \psi  + {w \mu \over \kappa}  (\bar q q)^2].
\label{n1mass}
\eea
Accordingly one has to add the following term in the fermionic supersymmetry transformation 
\bea
\delta'_\epsilon \psi= \sqrt 2 \mu \epsilon q, \quad 
\delta'_\epsilon \psi^\dagger = \sqrt 2 \mu \epsilon \bar q.
\label{n1masssusy}
\eea

\subsection{$\cN=2$} 
\label{n2}

$\cN=2$ superconformal Chern-Simons-matter theory with one chiral multiplet was studied in \cite{Gaiotto:2007qi}. The action with the gauge group $U(N)$ turns out to be obtained from $\cN=1$ superconformal action with the superpotential  \eqref{n1superpot} by setting $w=1$. 
For convenience we write down the explicit form of the action for $U(N)$ case.
\beal{
S^{\cN=2} 
&=\int d^3 x \biggl[ i\kappa \varepsilon^{\mu\nu\rho}\Tr (A_\mu\partial_\nu A_\rho -{2i\over3}  A_\mu  A_\nu A_\rho) + D_\mu \bar q D^\mu q + \bar\psi \sla D\psi \nn
& \qquad + {1 \over 2 \kappa} (\bar\psi q)( \bar q \psi)
+ {1 \over \kappa} (\bar\psi \psi) (\bar q q) 
+ \left( {1 \over 2 \kappa}\right)^2 (\bar q q)^3 \biggl].
\label{n2ukaction}
}

When the gauge group is $U(N)$, 
it is possible to add a mass term keeping $\cN=2$ supersymmetry, 
which is of the form \eqref{n1mass} with $w=1$.%
\footnote{In a case of $SU(N)$ gauge group there is no mass term preserving $\cN=2$ supersymmetry 
due to neither FI D-term nor gauge invariant superpotential.}
This is because in $U(N)$ case one can turn on an FI D-term, which generates a mass term 
by integrating out auxiliary adjoint fields.

\subsection{$\cN=3$} 
\label{n3action}

Let us consider $\cN=3$ $U(N)_k$ Chern-Simons-matter theory with minimal matter content, which is one fundamental hyper-multiplet. We denote two complex scalar by $q^A$ and its super-partners by $\psi_A$ in the hyper-multiplet, where $A=1,2$.
The action is given by
\beal{
S^{{\cal N}=3}  
&= \int d^3 x \biggl[  i\kappa \varepsilon^{\mu\nu\rho}\Tr (A_\mu\partial_\nu A_\rho -{2i\over3}  A_\mu  A_\nu A_\rho) +
D_\mu q^\dagger_A D^\mu q^A 
+ \psi^\dagger{}^{  A} \sla D \psi_{ A} \nn
&+ {1 \over  \kappa}(\psi^\dagger{}^{  A} \psi_B) (q^\dagger_A q^{  B}) 
+ {1 \over  \kappa}(\psi^\dagger{}^{  A} q^B) (q^\dagger_A \psi_{  B}) 
- {1 \over 2 \kappa} (\psi^\dagger{}^{  A} q^B) (q^\dagger_B \psi_{A}) \nn
&+ {1 \over 2 \kappa}  \varepsilon_{AB} \varepsilon_{CD} (\psi^\dagger{}^A q^B) 
(\psi^\dagger{}^C q^D) 
+ {1 \over 2\kappa}  \varepsilon^{AB} \varepsilon^{CD} ( q^\dagger_A\psi_B) (q^\dagger_C\psi_D) 
\nn
& + {1 \over \kappa^2}
\({1\over 3} (q_A^\dagger  q^B) (q^\dagger_C q^A) (q^\dagger_B q^C)
-{1\over 12} (q_A^\dagger  q^B) (q^\dagger_B q^C) (q^\dagger_C q^A)\)\biggl],
\label{n3ukaction}
}
where $\varepsilon_{12}=\varepsilon^{21}=1$ and the covariant derivative acts as \eqref{fcovderiv}.
Notice that the action has manifestly $SU(2)$ R-symmetry, which accounts for $\cN=3$ supersymmetry. 
The supersymmetry variation rule is given by
\bea
\delta_\omega q^A&=&-\sqrt2 \omega^{A  B}\psi_{  B},\quad
\delta_\omega q^\dagger_A=- \sqrt2\psi^\dagger{}^{  B} \omega_{AB},\\
\delta_\omega \psi_{  A}&=& 
\sqrt 2  (\omega_{CA} \sla D q^C+ {1 \over \kappa}\omega_{CB}(q^B
q^\dagger_A q^C -\half \delta_A^Bq^D q^\dagger_D q^C) ),\\
\delta_\omega \psi^\dagger{}^{  A}&=& 
\sqrt 2  (\omega^{CA} \sla D q^\dagger_C 
+ {1 \over \kappa}\omega^{CB}(q^\dagger_C q^A q^\dagger_B 
-\half \delta_B^A q^\dagger_C q^D q^\dagger_D)),\\
\delta_\omega A_\mu&=&{i \over \sqrt 2 \kappa}
\omega_{BA}\gamma^\mu (
q^A \psi^\dagger{}^B 
+ \varepsilon^{BC}\varepsilon^{DA} \psi_C q^\dagger_D),
\eea
where a supersymmetry parameter $\omega^{AB}$ is in the symmetric representation in $SU(2)$ R-symmetry: 
$ \varepsilon_{AB} \omega^{AB}=0$. We also use the following notation. 
\be
 \omega_{AB} := \varepsilon_{AC}\omega^{CD}\varepsilon_{DB} =
(\omega^{AB})^*  .
\ee

%%%%%%%%%%%%%%%%%%%%%%%%%%%%%%%%%%%%%%%
\subsection{$\cN=4$}
\label{n4}

$\cN=4$ Chern-Simons-matter theory with minimal matter content \cite{Gaiotto:2008sd} is given by a $U(N)_k\times U(N')_{-k}$
Chern-Simons theory with one bi-fundamental hyper-multiplet denoted by $q^A$ for two complex scalar and $\psi_A$ for their super-partners, where $A=1,2$.
The action is given by
\beal{
S^{{\cal N}=4}  
&= \int d^3 x \Tr \biggl[ i\kappa \varepsilon^{\mu\nu\rho} \left((A_\mu\partial_\nu A_\rho -{2i\over3}  A_\mu  A_\nu
A_\rho) -( A'_\mu\partial_\nu A'_\rho -{2i\over3}  A'_\mu  A'_\nu A'_\rho) \right)\nn
&+ D_\mu q^\dagger_A D^\mu q^A 
+ \psi^\dagger{}^{  A} \sla D \psi_{ A} \nn
&
+ {1 \over 2 \kappa} \left(q^\dagger_B q^B \psi^\dagger{}^{  A} \psi_{A}- q^Bq^\dagger_B \psi_{A}\psi^\dagger{}^{  A}
 -\varepsilon_{AC} \varepsilon_{BD} \psi^\dagger{}^A q^B \psi^\dagger{}^C q^D 
+\varepsilon^{AC} \varepsilon^{BD} \psi_A q^\dagger_B \psi_C q^\dagger_D \right) 
\nn
&+ {1 \over \kappa^2}
\biggl({3\over 2}  q_A^\dagger  q^B q^\dagger_B q^A q^\dagger_C q^C
-{2\over 3} q^A  q^\dagger_B q^C q^\dagger_A q^B q^\dagger_C 
-{5\over 12} (q^A q_A^\dagger  q^B  q^\dagger_B  q^C  q^\dagger_C
+q_A^\dagger  q^A q^\dagger_B q^B q^\dagger_C q^C ) \biggl)
\biggl].
\label{n4action}
}
The covariant derivative acts on the fields by \eqref{bifcovderiv}. 
This action has $SU(2)\times SU(2)$ R-symmetry, which explains $\cN=4$ supersymmetry. 
The supersymmetric transformation rule is given by
\beal{
\delta_\epsilon q^A&= -\sqrt 2 \epsilon^{A  B}\psi_{  B},\quad
\delta_\epsilon q^\dagger_A= - \sqrt 2 \psi^\dagger{}^{  B}\epsilon_{AB},\\
\delta_\epsilon \psi_{  A}
&= \sqrt 2  (\epsilon_{CA} \sla D  q^C+ {1 \over 2 \kappa} \epsilon_{CA} (q^C q^\dagger_D
q^D - q^D q^\dagger_D q^C) ),\\
\delta_\epsilon \psi^\dagger{}^{  A}
&= \sqrt 2  (\epsilon^{CA} \sla D  q^\dagger_C
+ {1 \over 2 \kappa} \epsilon^{CA} (q^\dagger_D q^D q^\dagger_C  
- q^\dagger_C q^D q^\dagger_D)),\\
\delta_\epsilon  A_\mu&=
{i \over \sqrt 2 \kappa}
\epsilon_{AB}\gamma^\mu (
q^A \psi^\dagger{}^B 
+ \varepsilon^{BC}\varepsilon^{DA} \psi_C q^\dagger_D), \\
\delta_\epsilon  A'_\mu&=
{i \over \sqrt 2 \kappa}
\epsilon_{AB}\gamma^\mu (
\psi^\dagger{}^B q^A + \varepsilon^{BC}\varepsilon^{DA}q^\dagger_D \psi_C ),
}
where $\epsilon^{AB}$ is a supersymmetry parameter with two independent $SU(2)$ indices and
\be
 \epsilon_{AB} := \varepsilon_{AC}\epsilon^{CD}\varepsilon_{DB} =
(\epsilon^{AB})^*.  
\ee

A mass term preserving not only $\cN=4$ but also $SO(4)_R$ symmetry was constructed in \cite{Hosomichi:2008jd}. In our notation, it is given by
\beal{
\cL^{\cN=4}_{\text{mass}}
&= \Tr \biggl[  \mu \psi^\dagger{}^{  A}\psi_{A}
+ \mu^2   q_A^\dagger q^A + {\mu \over \kappa} ( q^\dagger_A q^A q^\dagger_B q^B - q^\dagger_A q^B q^\dagger_B q^A) \biggl].
\label{n4mass}
}
Accordingly we add the following variation in the fermionic supersymmetry variation. 
\bea
\delta'_\epsilon \psi_{  A}= \sqrt 2  \mu \epsilon_{BA} q^B, \quad
\delta'_\epsilon \psi^\dagger{}^{  A}=  \sqrt 2  \mu \epsilon^{BA} q^\dagger_B.
\eea

\subsection{$\cN=6$} 
\label{n6}

We consider $\cN=6$ $U(N)_k\times U(N')_{-k}$ Chern-Simons 
theory with four complex bi-fundamental scalars denoted by $Y^A$ 
and its super-partner $\Psi_A$, where $A=1,2,3,4$ 
\cite{Aharony:2008ug,Aharony:2008gk}.%
\footnote{
In the terminology of superfield, the matter content of $\cN=6$ theory is one bi-fundamental
hyper-multiplet $(q^A, \psi_{\dot A})$ and anti-bi-fundamental (twisted) hyper-multiplet $(q^{\dot A}, \psi_{A})$.
The relation between these fields and $(Y^{\bA}, \Psi_{\bA})$ is given by
\beal{
Y^\bA =& (q^A, q^\dagger_{\dot A}), \quad Y^\dagger_\bA = (q^\dagger_A, q^{\dot A}) \nn
\Psi^\dagger{}^\bA =& (\varepsilon^{AB} \psi_B, \psi^\dagger{}^{\dot B}\varepsilon_{\dot B\dot A}), \quad 
\Psi_\bA = (\psi^\dagger{}^{B}\varepsilon_{BA}, \varepsilon^{\dot A\dot B} \psi_{\dot B}), \nn
\xi^{\bA\bB} =& 
\begin{pmatrix} 
 0& \epsilon^{A\dot C}\varepsilon_{\dot C\dot B} & \\ 
 \varepsilon^{BC} \epsilon_{C\dot A}& 0\\
\end{pmatrix}
}
where we use the notation $\bA, \bB$ representing $SU(4)$ indices only here. 
}
The action is given by
\beal{
S^{{\cal N}=6}  &= \int d^3 x \Tr \biggl[ i\kappa \varepsilon^{\mu\nu\rho} \left((A_\mu\partial_\nu A_\rho -{2i\over3} 
A_\mu  A_\nu A_\rho) -( A'_\mu\partial_\nu A'_\rho -{2i\over3}  A'_\mu  A'_\nu A'_\rho) \right) \nn 
&+ D_\mu Y^\dagger_A D^\mu Y^A + \Psi^\dagger{}^{  A} \sla D \Psi_{ A} \nn
&+ {1 \over 2 \kappa } ( Y^\dagger_A Y^A \Psi^\dagger{} ^B \Psi_B - Y^A Y^\dagger_A \Psi _B
\Psi^\dagger{} ^B+2 Y^A Y^\dagger_B \Psi _A \Psi^\dagger{} ^B - 2 Y^\dagger _A Y^B \Psi^\dagger{} ^A
\Psi_B \nn
&\quad -\varepsilon_{ABCD} \Psi^\dagger{}^A Y^B \Psi^\dagger{}^C Y^D 
+\varepsilon^{ABCD} \Psi_A Y^\dagger_B \Psi_C Y^\dagger_D ) \nn
&+{1 \over 12 \kappa^2 } \biggl( 
- Y^\dagger_A Y^A Y^\dagger_B Y^B Y^\dagger_C Y^C - Y^A Y^\dagger_A Y^B Y^\dagger_B Y^C Y^\dagger_C
- 4 Y^A Y^\dagger_B  Y^C  Y^\dagger_A Y^B Y^\dagger_C \nn
&\quad +6  Y^A Y^\dagger_B Y^B Y^\dagger_A Y^C Y^\dagger_C\biggl) \biggl].
\label{n6action}
}
Here $\varepsilon^{1234} = \varepsilon_{1234}=1$ and the covariant derivative acts on the fields by \eqref{bifcovderiv}. 
%\beal{
%D_\mu Y^A& = \partial_\mu Y^A -i A_\mu Y^A +i Y^A  A' {}_\mu, \quad
%D_\mu Y^\dagger_A= \partial_\mu Y^\dagger_A -i A' {}_\mu Y^\dagger_A +i Y^\dagger_A  A_\mu,\\
%D_\mu \Psi_A&= \partial_\mu\Psi_A -i A_\mu \Psi_A +i \Psi_A A' {}_\mu,\quad
%D_\mu \Psi^\dagger{}^{A}= \partial_\mu\Psi^\dagger{}^{A} -i A' {}_\mu \Psi^\dagger{}^{A}+i
%\Psi^\dagger{}^{A} A_\mu.
%}
Note that $SU(4)$ R-symmetry is explicitly seen and thus $\cN=6$ supersymmetry. 
The explicit supersymmetry variation rule is 
\bea
\delta_\xi Y^A &=&-\sqrt2 \xi^{AB} \Psi_B, \quad \delta_\xi Y^\dagger_A =-\sqrt2 \Psi^\dagger{}^{B} \xi_{AB},
\\
\delta_\xi \Psi_A&=&\sqrt2( \xi_{BA} \gamma^\mu D_\mu Y^B 
+\frac{1}{2\kappa } \xi_{CB} Q^B{}_A{}^C),
\label{dpsi2}
\\
\delta_\xi \Psi^\dagger{}^A&=&
\sqrt2(\xi^{BA} \gamma^\mu D_\mu Y^\dagger_B 
+ \frac{1}{2\kappa} \xi^{CB} (Q^B{}_A{}^C)^\dagger),
\label{dpsidag2}
\\
\delta_\xi A_\mu &=& \frac{i}{ \sqrt 2 \kappa} \left(  Y^B \Psi^\dagger{}^{ A} \gamma_\mu
\xi_{AB} - \xi^{AB} \gamma_\mu \Psi_B Y^\dagger_A \right), 
\\
\delta_\xi A' _\mu &=& \frac{ i }{ \sqrt2 \kappa} \left(  \Psi^\dagger{}^{A} \gamma_\mu \xi_{AB}
Y^B - Y^\dagger_A \xi^{AB} \gamma_\mu  \Psi_B \right),
\label{susyabjm}
\end{eqnarray}
where a supersymmetry parameter $\xi^{AB}$ is in the anti-symmetric representation in $SU(4)$ R-symmetry: 
$\xi^{AB}= -\xi^{BA}$. We also use the following notation. 
\bea
\xi_{AB}= -{1\over2}\varepsilon_{ABCD}\xi^{CD} = (\xi^{AB})^*
\label{xi}
\eea
%The bosonic potential term can be written as
%\be
%\cL_Y = \frac{1}{6 \kappa^2} Q^A{}_C{}^B (Q^A{}_C{}^B)^\dagger,
%\ee
and
\bea
Q^A{}_C{}^B =Y^A{}Y^\dagger_C{}Y^B + \delta_C^{A} Y^{[B}{}Y^\dagger_D{}Y^{D]}
-  \left( Y^B{}Y^\dagger_C{}Y^A + \delta_C^{B} Y^{[A}{}Y^\dagger_D{}Y^{D]} \right)
\eea
where the bracket means the normalized anti-symmetrization: $X^{[AB]}=\half (X^{AB} - X^{BA})$.

It is known that one can tern on a mass term keeping $\cN=6$ supersymmetry \cite{Gomis:2008vc}. 
The mass term is given by
\beal{
\cL^{\cN=6}_{\text{mass}}
&= \Tr \biggl[  \mu \psi^\dagger{}^{  A} M^B_A\psi_{  B}
+ \mu^2   Y_A^\dagger  M^A_B Y^B 
+ {\mu \over \kappa} M_A^B ( Y^\dagger_B Y^A Y^\dagger_C Y^C -Y^A Y^\dagger_B Y^C Y^\dagger_C )\biggl]
\label{n6mass}
}
where $M_{A}^{B} = \text{diag}(1,1,-1,-1)$.
The supersymmetry transformation is corrected so that one has to add the following variation in the fermionic
supersymmetry transformation rule. 
\bea
\delta'_\xi \psi_{  A}= \sqrt 2  \mu \xi_{CB} M^B_AY^C, \quad
\delta'_\xi \psi^\dagger{}^{  A}=  \sqrt 2  \mu \xi^{CB}  M^A_B Y^\dagger_C.
\eea

%%%%%%%%%%%%%%%%%%%%%%%%%%%%%%%%%%%%%%%%%%%
%\begin{thebibliography}{99}
%\end{thebibliography}
\bibliographystyle{utphys}
\bibliography{susycs}

\end{document}